\documentclass[11pt]{article}
\usepackage[margin=1in]{geometry}
\usepackage{setspace}
\usepackage{lineno}

\usepackage{subfig,multirow,listings,hyperref}
\usepackage{tcolorbox,amsmath}
\usepackage{amsmath}
\usepackage{graphicx,psfrag,epsf}
\usepackage{enumerate}
\usepackage{natbib}
\usepackage[ruled]{algorithm2e}
\usepackage{subfig}
\usepackage{url} 
\usepackage{multirow}
\usepackage{bm}
\RequirePackage[OT1]{fontenc}
\RequirePackage{amsthm,amsmath}
\usepackage{amsmath,amssymb}

\usepackage{microtype}

\begin{document}

\title{Sequential Bayesian Monitoring for Recoverable and Drifting Processes}

\author{
Gordon J. Ross\\
Department of Mathematics, University of Edinburgh, UK\\
\texttt{gordon.ross@ed.ac.uk}
}

\date{}


\maketitle

\begin{abstract}
In many Phase II statistical process control (SPC) problems, the main concern is not whether a monitored process has ever changed, but whether
it is currently operating at an acceptable level. This distinction is especially important when
monitoring continues after a signal, or when corrective action may restore the process. We develop Bayesian monitoring procedures for this
formulation of the Phase II task. For recoverable processes
that may alternate between in-control and out-of-control states, we derive
recursions for the posterior probability that the process is presently in
control. For sequential tracking problems in which a latent parameter
evolves over time, we monitor the posterior probability that the parameter lies
inside an acceptable region of behavior. The methods are
studied through calibrated time-between-failure experiments, Gaussian and
Binomial tracking examples, and a held-out multivariate data illustration using
white wine quality measurements.
\end{abstract}


\graphicspath{{img/}{../img/}}

\section{Introduction}

A common task within statistical process control (SPC) is to monitor a process in order to detect deviations away from a stable or acceptable state of operation. A typical example is the manufacture of polished wafers in the semiconductor industry \citep{Yeh2004,zhang_monitoring_2016}. The quality of a particular wafer depends on its thickness, with only a small range of values being desirable. Due to inherent statistical fluctuations in the manufacturing process, the thickness of individual wafers will vary, although the expected (i.e. population mean) thickness is constant when the process is operating as desired. However if a fault develops in the process then the expected wafer thickness may change, and this must be detected as soon as possible so that corrective action can be taken. A second classic example is a manufacturing process where each item has a certain probability of being defective. If a fault occurs, then the proportion of defective items may rise, and this again must be detected quickly to avoid financial loss \citep{Woodall1997,chukhrova_improved_2019}. Many more applications of SPC can be found in \cite{Montgomery2005} and \cite{Oakland2007}.

Suppose that a given process has $m$ quality characteristics of interest, such as the average wafer thickness or the probability of an item being defective. Let $\theta_t$ be a possibly vector-valued parameter denoting their values at each discrete time point $t  = 1,2,\ldots$. Typically $\theta_t$ is not directly observed, and instead one or more observations from the process are sampled at each time point. The observations at time $t$ are denoted by $y_t$ with distribution $y_t \sim f(y_t| \theta_t)$. Based on these observations, inference is made about $\theta_t$ and a warning is given if the process is found to have degraded in a way which has caused $\theta_t$ to move outside the range of values regarded as acceptable for the application. 

In this paper we focus on sequential process control, usually referred to as Phase II monitoring. In the standard Phase II setting, a reference sample of observations is first collected from the process during a period when it is believed to be operating in the desired in-control state. This is often called Phase I. The reference sample is used to learn about the in-control value of the process parameter, denoted here by $\theta^0$, and prospective monitoring then begins. The samples $y_1,y_2,\ldots$ are sequentially observed from the process, and if a fault occurs the observation distribution changes to $f(y_t | \theta^1)$, for some out-of-control parameter value $\theta^1 \neq \theta^0$. In the Bayesian formulation adopted here, a Phase I sample is not mathematically required: one could instead begin monitoring from a prior distribution for $\theta^0$. In many SPC applications, however, historical in-control data are available, and it is natural to use them to form a reference posterior distribution for the in-control state.

Traditional Phase II analysis is often presented as a one-way monitoring problem. The process is assumed to be in control until a signal is raised, at which point monitoring is stopped and the process is investigated or adjusted. This is appropriate in many industrial settings, but it is less suitable for applications where the process cannot simply be stopped. For example, \cite{Spiegelhalter2012} uses SPC ideas to monitor the performance of a group of hospitals. When the performance of a hospital is found to have deteriorated, it will not be practical to close it down. Instead, corrective action might be taken while the hospital remains open, and the relevant question is whether performance has subsequently returned to the desired level. After recovery, performance may degrade again, and so on. Similar settings have been considered by \cite{Gandy2013} and \cite{bodenham_continuous_2017}. In such cases, the objective is not only to detect whether a change has ever occurred, but to update the posterior probability that the process is currently in control after each new observation.

A related extension of SPC is \textbf{sequential tracking} \citep{Tsiamyrtzis2005,Tsiamyrtzis2008,Apley2012,wang_process_2018,hou_new_2020}. In this setting the process parameter may evolve over time, for example due to gradual equipment degradation, replacement of parts, or other natural variation in the process. Not every statistically detectable change is necessarily large enough to justify intervention. Instead, the practitioner may specify an acceptable region $A$ of the parameter space and seek to monitor the posterior probability that $\theta_t \in A$. Thus, while the recoverable-regime problem asks for the probability that the process is currently in an in-control state, the sequential-tracking problem asks whether a time-varying parameter remains inside a prespecified tolerance region.

A major motivation for Bayesian SPC is that uncertainty about the in-control parameter can be propagated directly through the monitoring procedure. This is important because replacing $\theta^0$ by a Phase I point estimate can substantially degrade the operating characteristics of control charts, particularly when the reference sample is short \citep{Jones2001,Jensen2006}. Related concerns have motivated a large literature on the effect of parameter estimation in control chart design, including work on guaranteed in-control performance and cautious parameter learning \citep{CapizziMasarotto2020}. Bayesian methods are attractive in this setting because the Phase I posterior can be used as the reference distribution for future monitoring, rather than treating the estimated in-control parameter as known.

There is now a substantial Bayesian SPC literature. Earlier Bayesian control charts include procedures for traditional Phase II monitoring and for posterior or posterior predictive monitoring of process parameters \citep{Bayarri2005,Tan2012,steward_bayesian_2016,Pan2017,Kumar2017,yazdi_new_2019,noor-ul-amin_adaptive_2021}. More recently, predictive Bayesian charts have been developed for short-run and self-starting settings. Predictive Control Charts (PCC) use posterior predictive distributions and possible historical information to monitor online processes \citep{Bourazas2022PCC}, while the Predictive Ratio CUSUM (PRC) provides a Bayesian memory-type chart for persistent shifts and has an associated design theory for choosing thresholds under false-alarm constraints \citep{Bourazas2023PRC,Bourazas2023PRCDesign}. Bayesian predictive approaches have also been developed specifically for time-between-events monitoring \citep{Kumar2017,Ali2020TBE}, and recent work has considered predictive Bayesian CUSUM charts under guaranteed in-control performance criteria \citep{WangCastagliolaGuo2026}. These methods are close in spirit to the present work, especially in their use of posterior predictive distributions and their concern with parameter uncertainty. The focus here is different: we model Phase II monitoring situations in which the process may leave and later return to the in-control state, and the central monitoring quantity is the posterior probability of being currently in control.

The computational structure of the proposed approach is also related to Bayesian online changepoint detection. In particular, online filtering methods can update posterior distributions over run lengths or changepoint locations as observations arrive \citep{Fearnhead2007,Chopin2007,AdamsMacKay2007}. However, the SPC formulation considered here has an additional state-space structure: one state represents the recurring in-control regime, while out-of-control episodes are treated as temporary departures from this reference behavior. This distinction matters because the inferential target is not simply the most recent changepoint, but the current in-control/out-of-control status of the process.

Our contribution is to formulate Phase II monitoring as Bayesian online inference on current process acceptability, with a recurring in-control reference state learned from Phase I information, temporary departures, and calibrated posterior signalling thresholds. The paper makes three main contributions. First, for recoverable processes that may alternate between in-control and out-of-control regimes, we derive exact recursive updates for the posterior distribution of the current regime and use $P(D_t=0\mid y_{1:t})$ as the monitoring statistic. Second, for sequential tracking problems in which the process parameter evolves over time, we formulate monitoring in terms of $P(\theta_t \in A\mid y_{1:t})$ for a prespecified acceptable region $A$, and describe particle-filter computation for nonlinear or multivariate cases where analytic filtering is unavailable. Third, we separate posterior inference from operating-threshold selection by calibrating posterior signalling thresholds to finite-horizon in-control behaviour under the Phase I reference distribution.

This last point is important for chart design. Although posterior probabilities provide interpretable monitoring statistics, a fixed posterior threshold does not automatically guarantee a desired in-control false-alarm behaviour. A posterior probability is an inferential quantity; an operating characteristic is induced only after a signalling rule has been chosen. In the operating-characteristic study below, the threshold is chosen by posterior-predictive calibration: conditional on the Phase I posterior, all-in-control future monitoring sequences are simulated, and the threshold is selected to give a desired finite-horizon false-signal rate. This calibration propagates uncertainty in the in-control parameter rather than treating the Phase I estimate as known.

The remainder of the paper proceeds as follows. Section 2 reviews the single-change-point Bayesian formulation, which provides useful background and notation. Section 3 develops the recoverable-regime model for processes that may leave and later return to the in-control state. Section 4 formulates sequential tracking over acceptable parameter regions, and Section 5 describes the use of Sequential Monte Carlo methods for computation in more general models. Section 6 studies operating characteristics for Exponential--Gamma time-between-failure monitoring, including posterior-predictive threshold calibration, comparison with a classical time-between-event chart, and sensitivity analyses for prior specification and Phase I information. Section 7 gives sequential-tracking examples, including a Kalman-filter validation and a nonlinear Binomial-logit tracking problem. Section 8 gives a held-out multivariate illustration using white wine quality measurements, and Section 9 concludes.

\section{Single-change Bayesian monitoring}
\label{sec:singleknownincontrol}

We first describe the standard Bayesian formulation for monitoring against a single persistent departure from the in-control state. This case is useful both as a point of comparison with existing Bayesian control charts and as a building block for the recoverable monitoring model developed in Section~\ref{sec:multiple}.

In traditional Phase II SPC the goal is to monitor a process to detect whether it has changed from the in-control state with observation distribution $f(y_t | \theta^0)$ to an out-of-control state with distribution $f(y_t|\theta^1)$. The observations $y_1,y_2,\ldots$ are received sequentially from the process, and it is assumed that at some point $\tau$ during this sequence the process may go out of control. Conditional on $\tau$ and on the model parameters, the sequence distribution is
\[
y_i \sim \left\{ \begin{array}{ll}
  f(y_i | \theta^0), & \mbox{if $i \leq \tau$,} \\
  f(y_i | \theta^1), & \mbox{if $i > \tau$, \quad $\theta^1 \neq \theta^0$.}
       \end{array} \right.
\]
The observations are typically assumed to be conditionally independent given $\tau$ and the relevant segment parameters. At each time point $t$, the observation $y_t$ is received and the monitoring procedure must decide whether there is sufficient evidence that a change has occurred. If a signal is raised, the process is flagged as out of control; otherwise monitoring continues with the next observation $y_{t+1}$.

In the Bayesian setting, priors are required for the changepoint and for the in-control and out-of-control parameters. We write $\tau \sim g(\cdot;\gamma)$, where $g(\cdot;\gamma)$ is a discrete prior on $\{1,2,3,\ldots\}$ with distribution function $G(\cdot;\gamma)$. A geometric prior gives a constant prior hazard for a change, while a negative binomial prior can be used to allow over-dispersion in segment lengths \citep{Fearnhead2006}. The duration prior describes prior beliefs about when a departure is likely to occur. It should not be confused with the calibration of the chart's false-alarm behaviour, which depends on the signalling threshold as well as on the monitoring statistic. For notational convenience, we suppress the dependence on $\gamma$ below.

Let $\pi_0$ denote the reference distribution for the in-control parameter at the start of Phase II monitoring. If a Phase I sample $x_1,\ldots,x_{n_I}$ is available from a period believed to be in control, then $\pi_0$ is the posterior distribution $p(\theta^0 | x_{1:n_I})$ obtained by updating an initial prior for $\theta^0$ using these data. If no Phase I sample is available, $\pi_0$ can instead be elicited directly from engineering or historical knowledge. The idealised case where $\theta^0$ is known exactly is recovered by taking $\pi_0$ to be a point mass. In all cases, $\pi_0$ is treated as the fixed reference distribution during Phase II monitoring; otherwise observations collected after monitoring begins could be allowed to redefine the in-control state.

Similarly, let $\pi_1$ denote the prior for the out-of-control parameter $\theta^1$. This prior specifies the departures from the in-control model that the chart is designed to detect efficiently. In some applications it may be natural to elicit $\pi_1$ directly; in others it may be specified through shifts relative to $\theta^0$, or as a mixture over several departures of practical concern. Since the out-of-control prior affects the power of the method against different alternatives, its role should be examined as part of the operating characteristics of the chart. We return to this point in the simulation study in Section~6.

It is useful to write the posterior distribution using predictive likelihoods. Let $m_0(y)$ denote the fixed in-control reference predictive density at the start of Phase II monitoring. If $\theta^0$ is known, then $m_0(y)=f(y\mid \theta^0)$. If $\theta^0$ is unknown and a Phase I sample has been observed, then
\[
  m_0(y)=\int f(y\mid \theta^0)\,\pi_0(d\theta^0),
\]
where $\pi_0$ is the Phase I posterior for $\theta^0$. This reference predictive distribution is held fixed during Phase II monitoring, so prospective observations are assessed for compatibility with the Phase I reference rather than being used to update the in-control state.

For the out-of-control parameter, define the segment marginal likelihood
\[
M_1(a,b)=\int \prod_{i=a}^{b} f(y_i\mid \theta^1)\,\pi_1(d\theta^1),
\]
with the convention that $M_1(a,b)=1$ when $a>b$. The corresponding in-control predictive likelihood for observations $y_a,\ldots,y_b$ is
\[
M_0(a,b)=\prod_{i=a}^{b} m_0(y_i),
\]
again taking $M_0(a,b)=1$ when $a>b$.

Next, associate a latent variable $C_t$ with each time point. Let $C_t=0$ if no change has occurred before time $t$, so that the process is still in control at time $t$. If the process went out of control after observation $j<t$, then set $C_t=j$ \citep{Fearnhead2007}. The posterior distribution of $C_t$ is then
\[
P(C_t =j | y_{1:t}) \propto \left\{ \begin{array}{ll}
\{1-G(t-1)\}M_0(1,t), & \mbox{if $j=0$,} \\
 g(j)M_0(1,j)M_1(j+1,t), & \mbox{if $1\leq j \leq t-1$.}
       \end{array} \right.
\]
The normalising constant is obtained by summing these unnormalised probabilities over $j=0,1,\ldots,t-1$. If the relevant priors are conjugate, the predictive and segment marginal likelihoods can often be evaluated analytically and updated recursively, making the computation suitable for sequential monitoring.

This posterior gives the probability that the process is still in control, $P(C_t=0 | y_{1:t})$, at each time point. We use this quantity as a posterior monitoring statistic and raise a signal when
\[
P(C_t=0 | y_{1:t}) < \delta,
\]
for some threshold $\delta$. A simple one-period generalised $0$--$1$ loss calculation motivates a rule of this form: if $cost_0$ is the cost of incorrectly signalling when the process is in control, and $cost_1$ is the cost of failing to signal when the process is out of control, then the one-period Bayes rule signals when
\begin{equation}
\delta  =  \frac{cost_1}{cost_0 + cost_1}.
\label{eqn:decision}
\end{equation}
However, a continuing control chart is a sequential decision problem, and the fully optimal policy would generally depend on delay costs and on the information expected from future observations. We therefore regard $\delta$ as an operating threshold for the posterior monitoring statistic. In Section~6, this threshold is chosen by posterior-predictive calibration to give a specified in-control false-signal behaviour.

The single-change formulation is limited because, after a departure occurs, the process is assumed either to remain out of control or the monitoring episode is effectively terminated. It does not represent settings where monitoring continues and the process may later return to the in-control state. The next section extends the latent-state formulation so that the target is the posterior probability that the process is currently in control, regardless of whether previous departures have occurred.

\section{Bayesian monitoring of recoverable regimes}
\label{sec:multiple}

Traditional Phase II SPC often assumes that, once the process has been flagged as out of control, monitoring is stopped while corrective action is taken. This is a reasonable approximation in many manufacturing settings, but it is not always appropriate. For example, \citet{Spiegelhalter2012} considers monitoring the performance of hospitals, where a deterioration in performance may trigger investigation or intervention but the hospital cannot simply be closed. Similar issues arise in financial risk monitoring \citep{Rosstechnometrics} and network intrusion monitoring \citep{Bodenham2013,bodenham_continuous_2017}. In such settings, monitoring continues after a signal, and the process may subsequently return to an acceptable state.

The relevant question is therefore not only whether a change has occurred at some time in the past, but whether the process is currently in control. We formulate this as a recoverable-regime problem: the process may alternate between an in-control regime and out-of-control regimes, with the in-control regime interpreted as a recurring reference state. Thus, at each time point $t$, the inferential target is
\[
P(D_t=0 \mid y_{1:t}),
\]
where $D_t=0$ denotes that the process is currently in control. This is a multiple-changepoint problem, but with an SPC-specific state structure. Rather than there being at most one changepoint, there may be changepoints $\tau_1,\tau_2,\ldots$ at which the process moves from in control to out of control or returns from out of control to in control:
\[
y_t \sim \left\{ \begin{array}{ll}
  f(y_t \mid \theta^0), & \mbox{if $t \leq \tau_1$,} \\
  f(y_t \mid \theta^1), & \mbox{if $\tau_1 < t \leq \tau_2$,} \\
  f(y_t \mid \theta^0), & \mbox{if $\tau_2 < t \leq \tau_3$,} \\
  f(y_t \mid \theta^2), & \mbox{if $\tau_3 < t \leq \tau_4$,} \\
  \ldots
       \end{array} \right.
\]
Here $\theta^0$ denotes the in-control reference parameter, while $\theta^1,\theta^2,\ldots$ denote the parameters associated with successive out-of-control episodes. These out-of-control parameters need not be equal, since different faults may produce different departures from the reference state.

The in-control regime is represented by a fixed reference distribution. This reference may be a point mass at a known value of $\theta^0$, but more typically it is the Phase I posterior distribution for $\theta^0$ or the associated posterior predictive distribution. Returning to the in-control state therefore means returning to compatibility with this reference distribution, rather than necessarily returning to a known numerical value of $\theta^0$. In the implementation considered here, the in-control reference is not updated using Phase II observations. This gives a stable monitoring reference: new observations are assessed for compatibility with the Phase I reference, rather than being allowed to redefine what counts as in control.

We focus on the case where departures are assessed relative to this recurring in-control reference, and where recovery corresponds to returning to it. More general models allowing direct transitions between distinct out-of-control regimes could be constructed, but would require additional transition and duration structure. They are not considered here. The filtering structure is related to Bayesian online changepoint methods such as \citet{Fearnhead2007}, but differs in making the current in-control/out-of-control state the primary inferential target. It also has similarities with alternating or epidemic changepoint models such as \citet{zhao_alternating_2021}, although the present formulation is motivated by Phase II monitoring.

Let $C_t$ denote the location of the most recent changepoint before time $t$. We use the convention that $C_t=r$ means that the current segment began with observation $y_{r+1}$. Thus $C_t=0$ if no changepoint has occurred and the current segment began at $y_1$. Let $D_t \in \{0,1\}$ denote the current regime, where $D_t=0$ is in control and $D_t=1$ is out of control. The process is assumed to begin in control, so states with $C_t=0$ and $D_t=1$ have zero prior mass. Define
\[
q_t(r,s)=P(C_t=r,D_t=s\mid y_{1:t}), \qquad r=0,\ldots,t-1,\quad s\in\{0,1\}.
\]
The posterior probability that the process is currently in control is then
\[
P(D_t=0\mid y_{1:t})=\sum_{r=0}^{t-1} q_t(r,0).
\]

Let $g_s(d)$ be the prior mass function for the duration of a segment in regime $s$, with distribution function $G_s(d)=\sum_{i=1}^d g_s(i)$, for $s\in\{0,1\}$. It is convenient to write the corresponding duration hazard as
\begin{equation}
 h_s(d)=\frac{g_s(d)}{1-G_s(d-1)}, \qquad d=1,2,\ldots,
\label{eqn:multipleprior}
\end{equation}
with $G_s(0)=0$. If the current segment is in regime $s$ and has lasted $d$ observations, then $h_s(d)$ is the prior probability that the segment ends after the current observation, while $1-h_s(d)$ is the prior probability that it continues.

Let $\pi_{\mathrm{OC}}$ denote the prior distribution for the parameter of a new out-of-control segment. This prior specifies the class of departures to which the chart is designed to be sensitive. Its choice is therefore part of the chart design, and its effect on operating characteristics is examined in Section~6. For an out-of-control segment beginning after observation $r$, let $\pi_{\mathrm{OC}}(\theta\mid y_{r+1:t})$ denote the posterior distribution for that segment parameter after observing the segment data $y_{r+1},\ldots,y_t$, with the convention that this posterior is just $\pi_{\mathrm{OC}}$ when $r=t$.

Now define the one-step predictive term
\begin{equation}
w^{(r,s)}_{t+1}=p(y_{t+1}\mid C_{t+1}=r,D_{t+1}=s,y_{1:t}).
\label{eqn:multiplepredictive}
\end{equation}
If $s=0$, this is the predictive density under the fixed in-control reference distribution. If $s=1$, it is the predictive density under the posterior for the current out-of-control segment,
\[
w^{(r,1)}_{t+1}=\int f(y_{t+1}\mid \theta)\,\pi_{\mathrm{OC}}(d\theta\mid y_{r+1:t}).
\]
When $r=t$, this is the prior predictive density for the first observation in a newly started out-of-control segment.

The filtering recursion can now be written compactly. Suppose $q_t(r,s)$ is available after observing $y_{1:t}$. For existing segments, $r<t$, the unnormalised posterior mass at time $t+1$ is
\begin{equation}
\tilde q_{t+1}(r,s)
=
w^{(r,s)}_{t+1}\{1-h_s(t-r)\}q_t(r,s),
\qquad r=0,\ldots,t-1,\quad s\in\{0,1\}.
\label{eqn:recoverable_continue}
\end{equation}
For a new changepoint after observation $t$, so that $C_{t+1}=t$, the two possible regime switches give
\begin{align}
\tilde q_{t+1}(t,0)
&=w^{(t,0)}_{t+1}
\sum_{i=0}^{t-1} h_1(t-i)q_t(i,1),
\label{eqn:recoverable_new_ic}\\
\tilde q_{t+1}(t,1)
&=w^{(t,1)}_{t+1}
\sum_{i=0}^{t-1} h_0(t-i)q_t(i,0).
\label{eqn:recoverable_new_ooc}
\end{align}
Finally, the posterior probabilities are obtained by normalising over all valid states,
\[
q_{t+1}(r,s)=\frac{\tilde q_{t+1}(r,s)}
{\sum_{j=0}^{t}\sum_{k=0}^{1}\tilde q_{t+1}(j,k)}.
\]
The desired posterior monitoring statistic is therefore
\[
p_{\mathrm{IC},t+1}=P(D_{t+1}=0\mid y_{1:t+1})=
\sum_{r=0}^{t}q_{t+1}(r,0).
\]
A signal is raised when $p_{\mathrm{IC},t}<\delta$. As in Section~\ref{sec:singleknownincontrol}, $\delta$ may be chosen from costs or expert tolerance, but in the simulation study below it is selected by posterior-predictive calibration to achieve a specified in-control false-signal rate.

The full procedure is summarized in Algorithm~\ref{fig:algorithm3}. The exact recursion requires updating $O(t)$ states when observation $y_{t+1}$ arrives, and hence has $O(T^2)$ computational cost over a monitoring horizon of length $T$. This is adequate for the moderate horizons used in Section~6. For very long streams, pruning or Sequential Monte Carlo approximations of the kind described in Section~\ref{sec:particlefilter} can be used to reduce the computational burden \citep{Fearnhead2007}.

\begin{algorithm}[t]
\footnotesize
Form the in-control reference distribution $\pi_0$ from Phase I data or prior information.\\
Choose duration distributions $g_0,g_1$, an out-of-control prior $\pi_{\mathrm{OC}}$, and a signalling threshold $\delta$.\\
After observing $y_1$, initialise $q_1(0,0)=1$ and $q_1(0,1)=0$.\\
\For{$t=1,2,\ldots$}{
Observe $y_{t+1}$.\\
Compute predictive terms $w^{(r,s)}_{t+1}$ for all valid states.\\
Update continuing-segment masses using Equation~\eqref{eqn:recoverable_continue}.\\
Update new-changepoint masses using Equations~\eqref{eqn:recoverable_new_ic}--\eqref{eqn:recoverable_new_ooc}.\\
Normalise to obtain $q_{t+1}(r,s)$.\\
Compute $p_{\mathrm{IC},t+1}=\sum_{r=0}^{t}q_{t+1}(r,0)$.\\
\eIf{$p_{\mathrm{IC},t+1}<\delta$}{
Flag that the process is currently out of control.
}{
Flag that the process is currently in control.
}
}
\caption{Bayesian Phase II monitoring algorithm for recoverable regimes}
\label{fig:algorithm3}
\end{algorithm}

The model is designed for settings in which departures are interpreted relative to a stable in-control reference, and recovery means return to compatibility with that reference. It is not intended to represent every possible form of nonstationary behaviour. Gradual drifts, transient outliers, contamination of Phase I data, and departures in directions not well represented by $\pi_{\mathrm{OC}}$ are considered empirically in Section~6 as operating-envelope checks.

\subsection{Example}
\label{sec:example}
We illustrate the filter using the example of monitoring the proportion of
defective items produced by a manufacturing process, a standard problem in Phase II SPC \citep{Montgomery2005,Rosscompstat}. The example is intended to show the posterior output of the recoverable-regime filter; formal threshold calibration and comparison with a classical method are given in Section~6.

At each time point, $N$ items are taken from the process and inspected, and the number of defectives is recorded. Let $y_t$ denote the number of defective items in the batch inspected at time $t$. Conditional on the current defective-item probability $\theta_t$, we assume
\[
y_t\mid \theta_t \sim \operatorname{Binomial}(N,\theta_t).
\]
If the in-control probability is not known exactly, a Phase I sample $x_1,\ldots,x_{n_I}$ can be used to form a Beta posterior reference distribution. For example, if the initial prior is $\operatorname{Beta}(a,b)$, then the Phase I posterior is $\operatorname{Beta}(a_0,b_0)$, where
\[
a_0=a+\sum_{i=1}^{n_I}x_i,
\qquad
b_0=b+n_I N-\sum_{i=1}^{n_I}x_i.
\]
For out-of-control segments, take $\pi_{\mathrm{OC}}=\operatorname{Beta}(a_1,b_1)$. The predictive terms in Equation~\eqref{eqn:multiplepredictive} are then
\[
w^{(r,s)}_{t+1}=\left\{\begin{array}{ll}
\binom{N}{y_{t+1}}(\theta^0)^{y_{t+1}}(1-\theta^0)^{N-y_{t+1}},
& \mbox{if $s=0$ and $\theta^0$ is known,}\\
\binom{N}{y_{t+1}}\dfrac{B(y_{t+1}+a_0,N-y_{t+1}+b_0)}{B(a_0,b_0)},
& \mbox{if $s=0$ and the Phase I reference is $\operatorname{Beta}(a_0,b_0)$,}\\
\binom{N}{y_{t+1}}\dfrac{B(y_{t+1}+\tilde a_1,N-y_{t+1}+\tilde b_1)}{B(\tilde a_1,\tilde b_1)},
& \mbox{if $s=1$,}
\end{array}\right.
\]
where
\[
\tilde a_1=a_1+\sum_{i=r+1}^{t}y_i,
\qquad
\tilde b_1=b_1+(t-r)N-\sum_{i=r+1}^{t}y_i.
\]
When $r=t$, the sums are empty and the out-of-control predictive distribution is the prior predictive distribution under $\operatorname{Beta}(a_1,b_1)$.

\begin{figure*}[t]
  \centering
  \subfloat[Process Data]{\label{fig:multipledata}\includegraphics[width=0.45\textwidth]{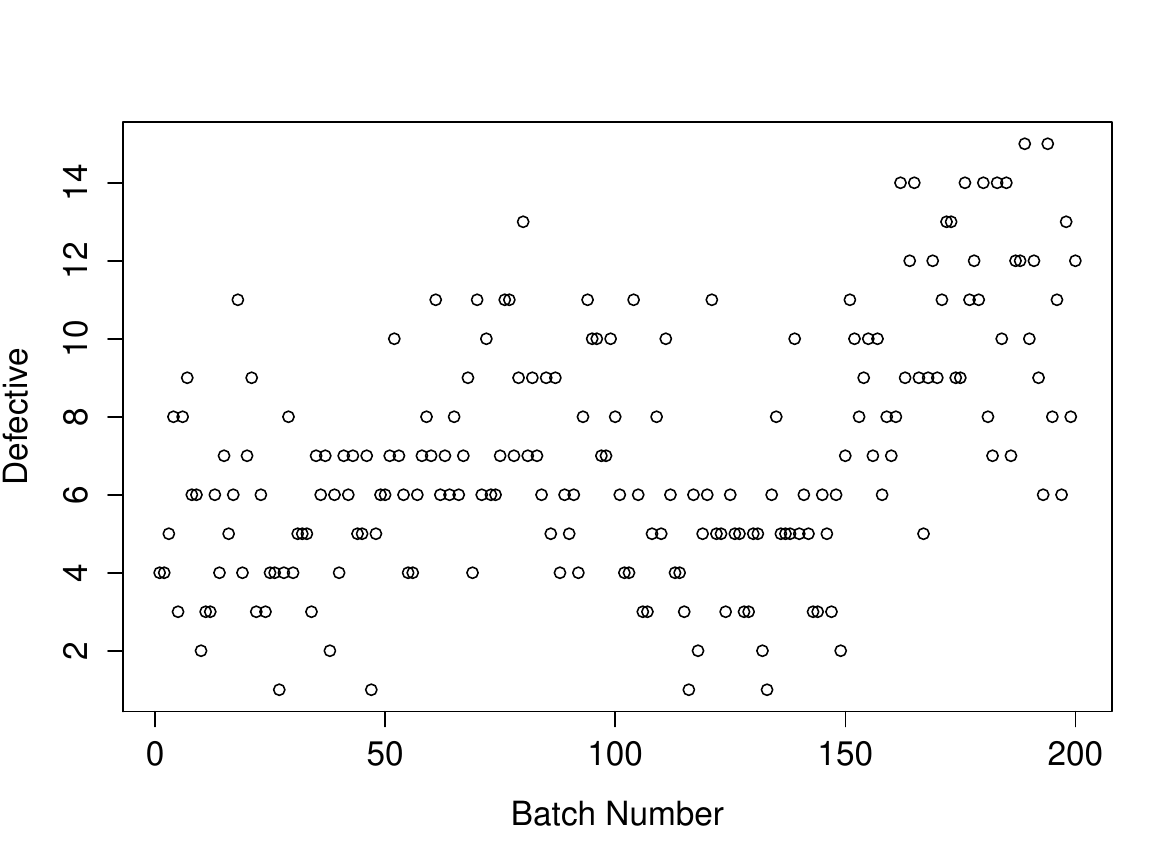}}
  \subfloat[In-control Probability]{\label{fig:multipleresults}\includegraphics[width=0.45\textwidth]{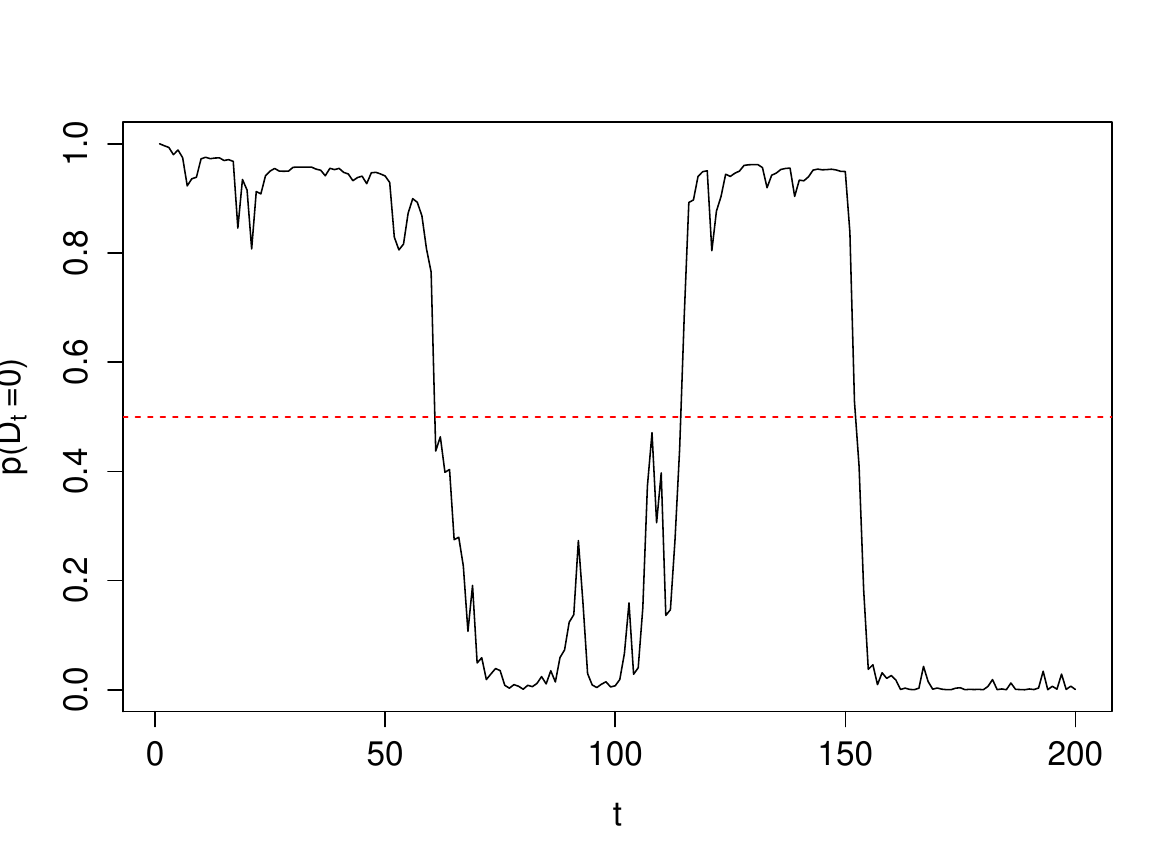}}
\caption{Sequential monitoring of a Bernoulli proportion that switches between the in-control state and two out-of-control episodes. The right panel shows the posterior probability $P(D_t=0\mid y_{1:t})$. The dotted line at $0.5$ is an illustrative threshold for this example; operating-characteristic-calibrated thresholds are considered in Section~6.}
  \label{fig:multiple}
\end{figure*}

We now simulate a single example sequence. When the process is in control, the proportion of defective items is $\theta^0=0.01$. The process goes out of control after time $50$, at which point the defective-item probability increases to $\theta^1=0.015$. It returns to the in-control state after time $100$, remains in control until time $150$, and then enters a second out-of-control episode with $\theta^2=0.02$. Monitoring stops at time $200$. At each time point $N=500$ items are inspected, and a simulated realisation is shown in Figure~\ref{fig:multipledata}.

For simplicity in this illustrative example, we take $\theta^0$ to be known and choose an out-of-control Beta prior with mean $0.02$ and standard deviation $0.01$. The in-control and out-of-control duration distributions are both taken to be geometric with mean $100$. Figure~\ref{fig:multipleresults} shows the resulting posterior in-control probability over time. Using the illustrative threshold $\delta=0.5$, the first out-of-control signal occurs at batch $t=61$, corresponding to a delay of $11$ batches. The signal persists until batch $t=115$, at which point the filter identifies that the process has returned to the in-control state. The second out-of-control signal occurs at batch $t=153$, where the larger change produces a shorter detection delay.

\section{Sequential tracking over acceptable regions}
\label{sec:tracking}

The recoverable-regime model in Section~\ref{sec:multiple} assumes that the process alternates between in-control and out-of-control regimes. A related but different Phase II problem arises when the process parameter itself evolves gradually over time. In this setting the aim is not necessarily to detect every departure from a fixed value, but to monitor whether the current parameter remains inside a prespecified acceptable region \citep{dette_detecting_2016}.

More formally, let $\theta_t$ denote a possibly vector-valued parameter containing the $m$ quality characteristics of interest at time $t$. The parameter is allowed to vary naturally over time, for example because of gradual wear, changes in operating conditions, or small adjustments to the process. This natural variation is not itself the target of intervention. Instead, a warning should be given if the current parameter value moves into a region regarded as unacceptable for the application. Let $A$ denote the subset of the parameter space corresponding to acceptable values. We say that the process is in control at time $t$ if $\theta_t \in A$, and out of control if $\theta_t \notin A$.

A simple choice is a rectangular tolerance region $A=\prod_{i=1}^m [L_i,U_i]$, where $[L_i,U_i]$ is the allowable range for the $i$th quality characteristic. This is not required, however. If the quality characteristics are dependent, or if acceptability is determined jointly rather than marginally, $A$ can instead be any prespecified joint region derived from engineering constraints, clinical tolerances, regulatory limits, or other subject-matter considerations. The set $A$ should also be distinguished from a control limit. It is a region of acceptable values for the latent process parameter, whereas the signalling threshold below is applied to a posterior monitoring statistic.

The natural time-variation in $\theta_t$ is modelled by a Markov transition density $p(\theta_t\mid\theta_{t-1};\gamma)$, where $\gamma$ denotes parameters governing the drift process. In what follows we condition on $\gamma$, treating it as specified from prior knowledge, estimated from historical data, or included in the state vector if sequential parameter learning is required. The transition density may include both gradual drift and abrupt jumps. As before, the observation $y_t$ is modelled through a measurement density $p(y_t\mid\theta_t;\gamma)$. This gives the state-space formulation
\begin{equation}
\begin{aligned}
\theta_t\mid\theta_{t-1} &\sim p(\theta_t\mid\theta_{t-1};\gamma),\\
y_t\mid \theta_t &\sim p(y_t\mid\theta_t;\gamma).
\end{aligned}
\label{eqn:statespace}
\end{equation}

We now give two simple examples which illustrate the formulation.

\textbf{Example 1: Monitoring a proportion.} Consider the case of monitoring the proportion of failures in a manufacturing process, as in Section~\ref{sec:example}. Let $\theta_t$ represent the proportion of faulty items at time $t$. If a batch of $N$ items is sampled at each time point, then the number of defective items $y_t$ follows a Binomial$(N,\theta_t)$ distribution. To allow $\theta_t$ to drift while preserving the constraint $0<\theta_t<1$, define
\[
z_t=\log\{\theta_t/(1-\theta_t)\}.
\]
A simple random-walk model on the logit scale is
\[
z_t=z_{t-1}+\varepsilon_t, \qquad \varepsilon_t\sim N(0,\sigma^2),
\]
with
\[
y_t\mid\theta_t \sim \mathrm{Binomial}(N,\theta_t).
\]
The drift parameter is then $\gamma=\{\sigma^2\}$, although abrupt jumps or heavier-tailed innovations could be incorporated if required. In a typical application, corrective action may only be needed if the proportion of defective items exceeds a tolerance level $U$, in which case $A=[0,U]$.

\textbf{Example 2: Gaussian drift and jumps.} The Bayesian SPC formulation considered by \cite{Tsiamyrtzis2005} and \cite{Tsiamyrtzis2010} concerns the case where $y_t$ has a univariate Gaussian distribution with mean $\theta_t$. In their work, the mean undergoes natural drift, represented by a random walk, and at each time step the process may jump by a fixed amount $\delta$. A simple version of this model is
\[
\theta_t\mid\theta_{t-1} \sim
\begin{cases}
N(\theta_{t-1},\sigma_\theta^2), & \mbox{with probability } 1-p,\\
N(\theta_{t-1}+\delta,\sigma_\theta^2), & \mbox{with probability } p,
\end{cases}
\]
with
\begin{equation}
y_t\mid\theta_t \sim N(\theta_t,\sigma_y^2).
\label{eqn:univariatemodel}
\end{equation}
The parameters associated with this process are $\gamma=\{p,\delta,\sigma_\theta^2,\sigma_y^2\}$, unless some are treated as fixed. The process is out of control at time $t$ if $\theta_t\notin [L,U]$ for a specified acceptable interval. Extensions with upward and downward jumps, or with random jump sizes, are considered by \cite{Tsiamyrtzis2010}.

In both examples, and in many similar models, the posterior distribution $p(\theta_t\mid y_{1:t})$ provides the basis for monitoring. The posterior probability that the process is currently in control is
\begin{equation}
p(\theta_t\in A\mid y_{1:t}) = \int_A p(\theta_t\mid y_{1:t})\,d\theta_t,
\label{eqn:tracking_probability}
\end{equation}
where the integral is over the acceptable region $A$. The sequence of probabilities in Equation~\ref{eqn:tracking_probability} can be used as a posterior monitoring statistic. A signal is raised when $p(\theta_t\in A\mid y_{1:t})<\delta$, where $\delta$ is an operating threshold chosen from costs, expert tolerance, or calibration to desired in-control behaviour.

The key implementation issue is the computation of the integral in Equation~\ref{eqn:tracking_probability}. When the transition and measurement densities in Equation~\ref{eqn:statespace} are linear and Gaussian, the posterior distribution of $\theta_t$ is available analytically from the Kalman filter \citep{Durbin2001}, and \cite{Tsiamyrtzis2005} show how this can be extended in simple jump models to give a tractable Gaussian-mixture posterior. In nonlinear or non-Gaussian settings, such as the Binomial-logit model above, these analytic calculations are no longer available.

Since Equation~\ref{eqn:statespace} is a state-space model, posterior quantities such as Equation~\ref{eqn:tracking_probability} can be approximated using sequential Monte Carlo methods. In \cite{Apley2012}, numerical quadrature was used to evaluate related posterior probabilities in Bayesian SPC. Sequential Monte Carlo provides a more general simulation-based alternative for nonlinear, non-Gaussian, and multivariate tracking problems; auxiliary, resample-move, Rao--Blackwellised, or other problem-specific particle-filter variants can be used when the basic bootstrap filter is not adequate.

\section{Sequential Monte Carlo computation}
\label{sec:smc}

The filters in Sections~\ref{sec:multiple} and~\ref{sec:tracking} are both sequential: after observing $y_t$, the posterior distribution needed for monitoring can be updated before $y_{t+1}$ arrives. In some conjugate recoverable-regime models the update can be performed exactly, as in Section~\ref{sec:multiple}. In more general state-space models, and in long recoverable-regime streams where retaining all possible changepoint states is undesirable, the same posterior quantities can be approximated using sequential Monte Carlo (SMC) methods, also known as particle filters. A general treatment of these methods is given by \citet{Doucet2001}; here we describe only the components needed for the monitoring procedures above.

\subsection{Particle filtering for sequential tracking}

In the sequential-tracking model of Section~\ref{sec:tracking}, the inferential target is the filtering distribution $p(\theta_t\mid y_{1:t})$ and, in particular, the posterior probability that the current parameter lies in the acceptable region $A$. A particle approximation represents this distribution by weighted samples
\[
  \left\{\theta_t^{(i)},w_t^{(i)}\right\}_{i=1}^P,
  \qquad \sum_{i=1}^P w_t^{(i)}=1.
\]
The posterior probability in Equation~\ref{eqn:tracking_probability} is then estimated by
\begin{equation}
  P(\theta_t\in A\mid y_{1:t})
  = \int_A p(\theta_t\mid y_{1:t})\,d\theta_t
  \approx \sum_{i=1}^P w_t^{(i)} I_A\{\theta_t^{(i)}\},
\label{eqn:tracking_particle_estimate}
\end{equation}
where $I_A\{\theta_t^{(i)}\}=1$ if the $i$th particle lies in $A$, and is zero otherwise.

The simplest implementation is the bootstrap, or sampling-importance-resampling, particle filter. Suppose that at time $t-1$ we have weighted particles approximating $p(\theta_{t-1}\mid y_{1:t-1})$. For each particle, draw a propagated value
\[
  \tilde\theta_t^{(i)} \sim p(\theta_t\mid \theta_{t-1}^{(i)}),
\]
using the transition density in Equation~\ref{eqn:statespace}. The particle is then weighted by the likelihood of the new observation,
\[
  \tilde w_t^{(i)} \propto w_{t-1}^{(i)} p(y_t\mid \tilde\theta_t^{(i)}),
\]
and the weights are normalised to sum to one. The monitoring probability $P(\theta_t\in A\mid y_{1:t})$ is estimated using Equation~\ref{eqn:tracking_particle_estimate}, and a signal is raised when this estimate falls below the chosen threshold $\delta$. To control weight degeneracy, the particles may then be resampled when the effective sample size falls below a prespecified threshold. The procedure is summarised in Algorithm~\ref{fig:algorithm4}.

\begin{algorithm}[h]
\footnotesize
Initialise particles $\theta_0^{(1)},\ldots,\theta_0^{(P)}$ from $p(\theta_0)$ and set $w_0^{(i)}=1/P$. Choose a signalling threshold $\delta$.\\
\For{each observation $y_t$}{
  \For{each $1\leq i\leq P$}{
    Draw $\tilde\theta_t^{(i)} \sim p(\theta_t\mid \theta_{t-1}^{(i)})$.\\
    Set $\tilde w_t^{(i)} = w_{t-1}^{(i)}p(y_t\mid \tilde\theta_t^{(i)})$.
  }
  Normalise the weights so that $\sum_{i=1}^P \tilde w_t^{(i)}=1$.\\
  Estimate $\hat p_{A,t}=\sum_{i=1}^P \tilde w_t^{(i)}I_A\{\tilde\theta_t^{(i)}\}$.\\
  \If{$\hat p_{A,t}<\delta$}{
    flag the process as currently out of control.
  }
  \eIf{the effective sample size is below the resampling threshold}{
    Resample indices $a_1,\ldots,a_P$ with probabilities $\tilde w_t^{(1)},\ldots,\tilde w_t^{(P)}$.\\
    Set $\theta_t^{(i)}=\tilde\theta_t^{(a_i)}$ and $w_t^{(i)}=1/P$ for $i=1,\ldots,P$.
  }{
    Set $\theta_t^{(i)}=\tilde\theta_t^{(i)}$ and $w_t^{(i)}=\tilde w_t^{(i)}$ for $i=1,\ldots,P$.
  }
}
\caption{Particle filtering for Bayesian sequential tracking}
\label{fig:algorithm4}
\end{algorithm}

The bootstrap filter is used here because it gives a transparent implementation of the posterior monitoring statistic. In applications where the transition model is diffuse, the observations are highly informative, or the dimension of the state vector is large, the same formulation can be combined with more efficient SMC variants, including auxiliary particle filters \citep{Pitt1999,Johansen2008}, stratified or systematic resampling, resample--move steps, or model-specific proposal distributions. Standard convergence results for particle filters imply that, under regularity conditions, weighted particle averages such as Equation~\ref{eqn:tracking_particle_estimate} converge to the corresponding posterior expectations as $P$ increases; see \citet{Crisan2002} for an overview.

If static parameters in the transition or observation model are uncertain, they can be estimated from historical data, assigned prior distributions and included in an enlarged particle state, or handled using particle-learning or resample--move extensions. The appropriate treatment is application dependent; the monitoring statistic itself remains the posterior probability of the acceptable region.

\subsection{Particle filtering for recoverable regimes}
\label{sec:particlefilter}

Sequential Monte Carlo can also be used to approximate the recoverable-regime filter in Section~\ref{sec:multiple}. The exact recursion in that section retains the full filtering distribution
\[
q_t(r,s)=P(C_t=r,D_t=s\mid y_{1:t}),
\]
over all possible most-recent changepoint locations $r=0,\ldots,t-1$ and regimes $s\in\{0,1\}$. This is feasible for moderate monitoring horizons and is the approach used in the simulation study below. For very long streams, however, retaining all possible values of $r$ may become computationally inconvenient. A particle approximation instead represents $q_t$ by weighted samples
\[
  \left\{(C_t^{(i)},D_t^{(i)}),w_t^{(i)}\right\}_{i=1}^P.
\]

Given particles at time $t$, propose a new latent state for each particle using the duration hazards in Equation~\ref{eqn:multipleprior}:
\[
  (\tilde C_{t+1}^{(i)},\tilde D_{t+1}^{(i)})
  \sim p(C_{t+1},D_{t+1}\mid C_t^{(i)},D_t^{(i)}).
\]
The proposed particle is then weighted by the predictive density of the next observation,
\[
  \tilde w_{t+1}^{(i)}
  \propto
  w_t^{(i)}
  p(y_{t+1}\mid \tilde C_{t+1}^{(i)},\tilde D_{t+1}^{(i)},y_{1:t}),
\]
where the predictive term is the corresponding quantity from Equation~\ref{eqn:multiplepredictive}. After normalisation, the current in-control probability is estimated by
\[
  \widehat P(D_{t+1}=0\mid y_{1:t+1})
  =
  \sum_{i=1}^P \tilde w_{t+1}^{(i)} I\{\tilde D_{t+1}^{(i)}=0\}.
\]
A signal is raised when this estimate is below the threshold $\delta$.

This particle approximation is directly analogous to the filtering method of \citet{Fearnhead2007} for online multiple-changepoint problems, but with the additional regime indicator $D_t$ used to distinguish in-control and out-of-control states. Since $(C_t,D_t)$ is discrete, stratified or systematic resampling can be used to improve efficiency. In the Exponential--Gamma experiments in Section \ref{sec:simulation} we use the exact recursion rather than this approximation, because the monitoring horizon is short and exact filtering avoids Monte Carlo error.

\section{Simulation Study}
\label{sec:simulation}

This section studies the operating behaviour of the recoverable-regime methodology from Section~\ref{sec:multiple} in a time-between-failure setting. The focus is the setting for which the current-status formulation is designed: monitoring continues after a signal, the process may return to the in-control state, and subsequent out-of-control episodes may occur. We examine detection delay, recovery delay, false-signal behaviour, threshold calibration, and sensitivity to prior specification and Phase I information. These operating characteristics are central to the proposed formulation because the Bayesian procedure depends on both the in-control reference distribution and the prior used for out-of-control segment parameters.

Throughout this section the observations are time-between-failure values. Conditional on the current process rate $\theta_t$, we assume
\[
  Y_t \mid \theta_t \sim \operatorname{Exponential}(\theta_t),
\]
where the exponential distribution is parameterised by its rate. The in-control rate is $\theta^0=10$. Larger values of $\theta_t$ correspond to shorter expected times between failures and hence to process degradation. Gamma priors are used for the unknown rates, with shape--rate parameterisation. When a Gamma prior is specified through a mean $\mu$ and standard deviation $\sigma$, the corresponding shape and rate are
\[
  \alpha = \frac{\mu^2}{\sigma^2}, \qquad
  \beta = \frac{\mu}{\sigma^2}.
\]
For the in-control rate, a Phase I sample $X_1,\ldots,X_m$ is generated from the in-control process and used to update the prior for $\theta^0$. If the prior is $\operatorname{Gamma}(\alpha,\beta)$, the Phase I posterior is
\[
  \theta^0 \mid X_{1:m} \sim
  \operatorname{Gamma}\left(\alpha+m,\,\beta+\sum_{i=1}^m X_i\right).
\]
This posterior is then used as the fixed in-control reference distribution during Phase II monitoring. Each out-of-control segment has a new rate parameter with prior $\operatorname{Gamma}(\alpha_1,\beta_1)$. The duration distributions for both in-control and out-of-control segments are geometric with parameter $1/200$.

\subsection{Posterior-predictive threshold calibration}
\label{subsec:threshold-calibration}

The monitoring statistic for the Bayesian procedure is the posterior probability
\[
  p_t = p(D_t=0 \mid y_{1:t}),
\]
where $D_t=0$ denotes the event that the process is currently in control. A signal is raised when $p_t < \delta$. Rather than fixing $\delta$ arbitrarily, we calibrate it by posterior prediction under in-control operation.

For a fixed Phase I data set in an applied problem, this calibration would be performed conditionally on the observed Phase I posterior. In the simulation study, however, the Phase I sample is itself random. We therefore use a pre-posterior version of the same calibration. For each calibration replicate, a Phase I sample is generated from the in-control process, the corresponding posterior for $\theta^0$ is obtained, and an all-in-control Phase II sequence of length 200 is generated from the posterior predictive distribution. That is, a rate is drawn from the Phase I posterior and a future sequence is then simulated conditional on this draw. The Bayesian filter is run once on this sequence, producing the path $p_1,\ldots,p_{200}$. Candidate thresholds are evaluated from this stored path, so the filter is not rerun separately for each value of $\delta$.

A false-signal episode is defined as a transition from no signal to signal, treating the process as non-signalling before the first observation. The selected threshold is the value of $\delta$ for which the expected number of false-signal episodes in 200 all-in-control observations is closest to one. This is a finite-horizon analogue of calibrating a chart to an in-control average run length of approximately 200, but it propagates uncertainty in $\theta^0$ rather than replacing it by a point estimate.

For the baseline configuration used below, $\mu_0=10$, $\sigma_0=3$, $m=50$, $\mu_1=40$, and $\sigma_1=10$. Using $1000$ posterior-predictive calibration sequences and a grid $\delta\in\{0.005,0.010,\ldots,0.995\}$, the calibrated threshold was
\[
  \delta = 0.485.
\]
The achieved mean number of false-signal episodes was $0.992$ with Monte Carlo standard error $0.049$. In the earlier version of this paper a fixed threshold was used; the results below use calibrated thresholds unless otherwise stated.

\subsection{Baseline comparison with a classical time-between-event chart}
\label{subsec:baseline-comparison}

We first compare the Bayesian filter with the classical time-between-event change detection method of \cite{RossSC}. We use the \texttt{ExponentialAdjusted} CPM with nominal $ARL_0=200$ and startup value 20. This is a fairer comparator than a chart given the true value of $\theta^0$, since it does not require the in-control rate to be known exactly. The two procedures do not have identical calibration mechanisms: the CPM is specified through a nominal $ARL_0$, whereas the Bayesian threshold is chosen to give one expected false-signal episode in 200 posterior-predictive in-control observations. We use these settings as comparable finite-horizon false-signal targets rather than as exactly equivalent operating constraints.

In the multiple-change setting we use the standard streaming version of the CPM, which continues after a detected change by reinitialising the chart after the estimated changepoint. This reset rule differs from the Bayesian procedure, which directly estimates the current in-control probability at every time point, so comparisons should be interpreted in terms of detected signal episodes rather than time spent in a signalled state.

Two benchmark scenarios are considered. The first is a standard single-change setting,
\[
Y_t \sim
\begin{cases}
  \operatorname{Exponential}(10), & 1\leq t\leq 100,\\
  \operatorname{Exponential}(40), & 101\leq t\leq 200.
\end{cases}
\]
The second is the recoverable setting,
\[
Y_t \sim
\begin{cases}
  \operatorname{Exponential}(10), & 1\leq t\leq 50,\\
  \operatorname{Exponential}(40), & 51\leq t\leq 100,\\
  \operatorname{Exponential}(10), & 101\leq t\leq 150,\\
  \operatorname{Exponential}(50), & 151\leq t\leq 200.
\end{cases}
\]
For the single-change scenario, $d_1$ is the first signal time after the true changepoint minus 100. For the recoverable scenario, $d_1$ is the first signal in observations 51--100 minus 50, $d_2$ is the first return to non-signal in observations 101--150 minus 100 for the Bayesian method, and $d_3$ is the first signal in observations 151--200 minus 150. For the CPM, which emits changepoint detections rather than a current in-control probability, $d_2$ is defined as the first detected change in observations 101--150 after reset. False-signal episodes are counted by the time at which an episode begins; a signal episode which begins in a true out-of-control segment and persists into a subsequent in-control segment is treated as recovery delay rather than as a false positive. Each row in Table~\ref{tab:baseline-comparison} is based on 1000 simulated sequences. If a detection or recovery event is not observed in the relevant segment, its delay is treated as missing; reported delay means are therefore conditional on the event being observed, and miss rates are reported separately. Monte Carlo standard errors are shown in parentheses.

\begin{table}[tbp]
\centering
\scriptsize
\setlength{\tabcolsep}{4pt}
\begin{tabular*}{\hsize}{@{\extracolsep{\fill}}llccccc@{}}
\hline
Scenario & Method & $d_1$ & $d_2$ & $d_3$ & Miss & $F$ \\
\hline
Single change & Bayesian & 6.49 (0.09) & -- & -- & 0.000 & 0.44 (0.03) \\
Single change & CPM & 8.20 (0.27) & -- & -- & 0.014 & 0.62 (0.03) \\
Recoverable & Bayesian & 6.57 (0.09) & 4.56 (0.11) & 5.66 (0.06) & 0.000/0.000/0.000 & 0.54 (0.03) \\
Recoverable & CPM & 7.70 (0.17) & 5.55 (0.15) & 6.33 (0.13) & 0.012/0.007/0.014 & 0.37 (0.02) \\
\hline
\end{tabular*}
\caption{Baseline comparison between the Bayesian recoverable-regime filter and the exponential CPM comparator. The Bayesian method uses the posterior-predictive calibrated threshold $\delta=0.485$. Delay entries are means over 1000 simulated sequences, with Monte Carlo standard errors in parentheses. The column $F$ gives the mean number of false-signal episodes. Miss rates are reported as $d_1/d_2/d_3$ for the recoverable scenario.}
\label{tab:baseline-comparison}
\end{table}

The Bayesian procedure detects the out-of-control changes faster than the CPM in both scenarios. In the single-change scenario the mean delay is 6.49 observations for the Bayesian method compared with 8.20 for the CPM. In the recoverable scenario the Bayesian method also has shorter mean delays for both out-of-control episodes and for recovery. The CPM has fewer false-signal episodes in the recoverable setting, but more false-signal episodes in the single-change setting. The comparison is best interpreted as an operating-characteristic comparison between two different monitoring summaries: the Bayesian method produces a persistent posterior probability of current in-control status, whereas the CPM produces discrete changepoint detections. After posterior-predictive calibration, the Bayesian filter gives competitive false-signal behaviour while providing faster detection and direct posterior information about the current process state.

\subsection{Sensitivity to prior specification and Phase I information}
\label{subsec:prior-sensitivity}

We next examine how the Bayesian procedure is affected by the prior distributions and the amount of Phase I information. The recoverable data-generating process is the same as in the second benchmark above. For each prior configuration, a new threshold is selected by posterior-predictive calibration using 500 all-in-control calibration sequences, again targeting one false-signal episode in 200 observations. The exception is the baseline row, where we use the more precise calibration from Section~\ref{subsec:threshold-calibration} based on 1000 calibration sequences. The evaluation results are based on 1000 simulated recoverable sequences. Table~\ref{tab:prior-sensitivity} reports the calibrated threshold, the achieved calibration false-signal rate, the three delay measures, false-signal episodes, and miss rates. Delay means are computed over non-missing detections or recoveries, with miss rates reported separately. Monte Carlo standard errors are shown in parentheses for the delay and false-episode means.

\begin{table}[tbp]
\centering
\scriptsize
\setlength{\tabcolsep}{4pt}
\begin{tabular*}{\hsize}{@{\extracolsep{\fill}}rrrrrrrr@{}}
\hline
Row & $\mu_0$ & $\sigma_0$ & $m$ & $\mu_1$ & $\sigma_1$ & $\delta$ & $F_{\rm cal}$ \\
\hline
1 & 10 & 3 & 50 & 40 & 10 & 0.485 & 0.99 \\
2 & 10 & 3 & 50 & 20 & 10 & 0.225 & 0.99 \\
3 & 10 & 3 & 50 & 100 & 10 & 0.650 & 1.01 \\
4 & 10 & 3 & 50 & 60 & 10 & 0.605 & 1.00 \\
5 & 15 & 5 & 50 & 40 & 10 & 0.470 & 1.00 \\
6 & 15 & 5 & 5 & 40 & 10 & 0.035 & 1.00 \\
7 & 1000 & 10000 & 50 & 40 & 10 & 0.475 & 1.00 \\
8 & 1000 & 10000 & 5 & 40 & 10 & 0.005 & 0.68 \\
\hline
\multicolumn{8}{c}{}\\[-1.5ex]
\hline
Row & \multicolumn{2}{c}{$d_1$} & \multicolumn{2}{c}{$d_2$} & \multicolumn{2}{c}{$d_3$} & $F$ \\
\hline
1 & \multicolumn{2}{c}{6.57 (0.09)} & \multicolumn{2}{c}{4.56 (0.11)} & \multicolumn{2}{c}{5.66 (0.06)} & 0.54 (0.03) \\
2 & \multicolumn{2}{c}{9.69 (0.11)} & \multicolumn{2}{c}{3.84 (0.09)} & \multicolumn{2}{c}{8.37 (0.09)} & 0.37 (0.02) \\
3 & \multicolumn{2}{c}{7.80 (0.16)} & \multicolumn{2}{c}{2.26 (0.05)} & \multicolumn{2}{c}{5.59 (0.11)} & 0.47 (0.02) \\
4 & \multicolumn{2}{c}{6.43 (0.11)} & \multicolumn{2}{c}{3.80 (0.09)} & \multicolumn{2}{c}{5.14 (0.07)} & 0.64 (0.03) \\
5 & \multicolumn{2}{c}{7.21 (0.10)} & \multicolumn{2}{c}{4.27 (0.10)} & \multicolumn{2}{c}{5.89 (0.07)} & 0.45 (0.02) \\
6 & \multicolumn{2}{c}{17.23 (0.24)} & \multicolumn{2}{c}{1.88 (0.04)} & \multicolumn{2}{c}{13.52 (0.15)} & 0.16 (0.01) \\
7 & \multicolumn{2}{c}{6.83 (0.10)} & \multicolumn{2}{c}{4.50 (0.10)} & \multicolumn{2}{c}{5.77 (0.07)} & 0.56 (0.03) \\
8 & \multicolumn{2}{c}{16.96 (0.35)} & \multicolumn{2}{c}{1.28 (0.02)} & \multicolumn{2}{c}{15.25 (0.30)} & 0.47 (0.06) \\
\hline
\multicolumn{8}{c}{}\\[-1.5ex]
\hline
Row & \multicolumn{7}{c}{Miss rates for $d_1/d_2/d_3$} \\
\hline
1 & \multicolumn{7}{c}{0.000/0.000/0.000} \\
2 & \multicolumn{7}{c}{0.000/0.000/0.000} \\
3 & \multicolumn{7}{c}{0.001/0.000/0.000} \\
4 & \multicolumn{7}{c}{0.000/0.000/0.000} \\
5 & \multicolumn{7}{c}{0.000/0.000/0.000} \\
6 & \multicolumn{7}{c}{0.004/0.000/0.000} \\
7 & \multicolumn{7}{c}{0.000/0.000/0.000} \\
8 & \multicolumn{7}{c}{0.189/0.000/0.113} \\
\hline
\end{tabular*}
\caption{Sensitivity of the Bayesian recoverable-regime filter to prior specification and Phase I sample size. The first panel gives the prior settings and calibrated thresholds, where $F_{\rm cal}$ is the achieved mean number of false-signal episodes in the all-in-control calibration experiment. The second panel gives evaluation performance over 1000 recoverable sequences; delay entries are means with Monte Carlo standard errors in parentheses, and $F$ is the mean number of false-signal episodes. Row 1 uses the full baseline calibration with $B_{\rm cal}=1000$ from Section~\ref{subsec:threshold-calibration}; rows 2--8 use independent row-specific calibrations with $B_{\rm cal}=500$. Row 8 is a stress case: the selected threshold was the smallest grid value, and the calibration did not cleanly attain the target false-signal rate.}
\label{tab:prior-sensitivity}
\end{table}

Several patterns are apparent. Moderate misspecification of the in-control prior has only a limited effect when the Phase I sample is reasonably large. For example, row 5 uses an in-control prior centred at 15 rather than 10, but with $m=50$ its performance is close to the baseline. Similarly, row 7 uses a very diffuse in-control prior, but with $m=50$ the Phase I sample is sufficient to recover baseline-like performance. In contrast, a short Phase I sample can be damaging when the prior is also inaccurate: row 6 has substantially longer detection delays for both out-of-control episodes.

The out-of-control prior also matters. When the prior for out-of-control rates is centred too low, as in row 2, detection of the two degraded periods is slower. When the prior is centred too high, as in row 3, the first out-of-control detection is also slowed and there is a small miss rate, although recovery is detected quickly. The final row gives an intentionally severe stress case with a vague in-control prior and only five Phase I observations. In this case the calibration curve is unstable, the calibrated threshold is extremely small, and the method has substantial miss rates for both out-of-control episodes. The calibration target is difficult to attain on the chosen threshold grid, reflecting the weak in-control reference produced by combining a vague prior with only five Phase I observations. These results illustrate a basic feature of the Bayesian formulation: an informative prior can improve detection when it reflects the departures of practical concern, but it should not be interpreted as a universally robust default. Prior specification is part of the chart design.

\subsection{Operating-envelope checks}
\label{subsec:operating-envelope}

Finally, we examine three types of departure from the baseline assumptions. These are not intended as new benchmark comparisons, but as checks on the operating envelope of the Bayesian procedure. The baseline prior configuration is used throughout, and the threshold is fixed at the baseline calibrated value $\delta=0.485$ rather than recalibrated for each stress case.

The first type of departure is contamination of the Phase I sample. Each Phase I observation is drawn from the in-control distribution with probability $1-\epsilon$ and from $\operatorname{Exponential}(40)$ with probability $\epsilon$, with $\epsilon=0.05$ or $0.10$. The Phase II process is otherwise the recoverable benchmark. The second departure replaces the first abrupt out-of-control transition by a gradual ramp, with the rate increasing linearly from 10 to 40 over observations 51--100. The third departure keeps the out-of-control prior centred on rate increases near 40, but the true out-of-control rate is 5 in both out-of-control segments. This represents a departure in the opposite direction from the prior expectation.

\begin{table}[tbp]
\centering
\tiny
\setlength{\tabcolsep}{2pt}
\begin{tabular*}{\hsize}{@{\extracolsep{\fill}}lccccc@{}}
\hline
Scenario & $d_1$ & $d_2$ & $d_3$ & Miss & $F$ \\
\hline
Clean baseline & 6.57 (0.09) & 4.56 (0.11) & 5.66 (0.06) & 0.000/0.000/0.000 & 0.54 (0.03) \\
5\% Phase I contamination & 6.92 (0.10) & 4.37 (0.10) & 5.88 (0.07) & 0.000/0.000/0.000 & 0.49 (0.03) \\
10\% Phase I contamination & 7.23 (0.10) & 4.30 (0.10) & 6.09 (0.07) & 0.000/0.000/0.000 & 0.49 (0.03) \\
Ramp first OOC & 22.65 (0.28) & 6.12 (0.15) & 5.62 (0.07) & 0.001/0.001/0.000 & 0.71 (0.03) \\
Wrong-direction OOC & 37.21 (0.66) & 15.34 (0.88) & 20.76 (0.66) & 0.769/0.078/0.444 & 0.60 (0.03) \\
\hline
\end{tabular*}
\caption{Operating-envelope checks for the Bayesian recoverable-regime filter. The baseline calibrated threshold $\delta=0.485$ is used for all rows. The clean baseline row is repeated from Table~\ref{tab:baseline-comparison}. The ramp delay $d_1$ is measured from the start of the ramp at $t=50$, so it should not be interpreted in the same way as a step-change delay. The column $F$ gives the mean number of false-signal episodes.}
\label{tab:operating-envelope}
\end{table}

The contamination scenarios are less damaging than might be expected. With 5\% contamination, the mean posterior Phase I reference rate increases to 10.48; with 10\% contamination it increases to 10.77. Nevertheless, the detection delays increase only modestly and no misses occur in these experiments. This should not be read as a general robustness guarantee against arbitrary Phase I contamination, but it suggests that mild contamination does not destroy performance in this setting.

The ramp scenario is more difficult for the first out-of-control episode, as expected. The mean first signal time is 72.65, which is 22.65 observations after the beginning of the ramp. However, the rate first exceeds 20 at observation 67 and first exceeds 30 at observation 84, so the mean signal occurs about 5.65 observations after the rate exceeds 20 and before the rate reaches 30. Thus the method is slower relative to the nominal start of the segment, but it still reacts while degradation is developing. Recovery and the second abrupt out-of-control episode remain close to the baseline.

The wrong-direction scenario is the clearest failure mode. The prior for out-of-control segments is centred on increased failure rates, while the true out-of-control rate is lower than the in-control rate. In this case the method misses the first out-of-control episode in 76.9\% of replicates and the second in 44.4\%, so the reported delay means for this row should be interpreted as conditional on the relatively small subset of runs where a signal is eventually raised. This behaviour is consistent with the role of the out-of-control prior: the filter has been designed to detect degradation in the direction encoded by that prior. In applications where both degradation and improvement are relevant, the prior for out-of-control episodes should be broadened or constructed as a mixture over the departures of interest.

Overall, the simulation study supports three conclusions. First, posterior-predictive calibration provides a transparent way to choose the Bayesian signalling threshold while accounting for Phase I uncertainty. Second, in the recoverable time-between-failure setting, the Bayesian filter gives competitive false-signal behaviour and shorter mean detection/recovery delays than the classical CPM comparator. Third, the method is sensitive to the prior placed on out-of-control segment parameters and to the amount of Phase I information available. This sensitivity reflects the modelling problem rather than the numerical computation: an informed Bayesian chart is most useful when the prior represents the departures that the practitioner wants to detect.

\section{Sequential Tracking Examples}
\label{sec:tracking-examples}

Section~\ref{sec:simulation} focused on the exact recoverable-regime filter from Section~\ref{sec:multiple}. We now illustrate the sequential-tracking formulation from Section~\ref{sec:tracking}, where the target is the posterior probability that a time-varying parameter remains inside a prespecified acceptable region. The first example is a Gaussian model for which the exact filtering distribution is available from the Kalman filter, and is used to validate the particle-filter approximation. The second example considers a nonlinear Binomial-logit model for monitoring a drifting defect probability.

\subsection{Gaussian tracking: comparison with the Kalman filter}
\label{subsec:gaussian-tracking}

We first consider a univariate Gaussian tracking problem. This example is deliberately chosen so that the exact posterior distribution is available analytically, allowing the particle filter to be checked against the Kalman filter.

The latent process is a deterministic drift-and-recovery trajectory of length $T=200$:
\[
\theta_t =
\begin{cases}
0, & 1 \leq t \leq 50,\\
0.9(t-50)/50, & 51 \leq t \leq 100,\\
0.9, & 101 \leq t \leq 140,\\
0.9\{1-(t-140)/40\}, & 141 \leq t \leq 180,\\
0, & 181 \leq t \leq 200.
\end{cases}
\]
Observations are generated from
\[
  Y_t\mid \theta_t \sim N(\theta_t,\sigma_y^2),
  \qquad \sigma_y=0.15.
\]
The acceptable region is
\[
  A=[-0.5,0.5].
\]
Under this construction, the latent process first leaves the acceptable region at $t=78$ and returns to it at $t=158$.

The filtering model used by both the Kalman filter and the particle filter is the Gaussian random walk
\[
  \theta_t\mid \theta_{t-1}\sim N(\theta_{t-1},\sigma_\theta^2),
  \qquad
  Y_t\mid \theta_t\sim N(\theta_t,\sigma_y^2),
\]
with $\sigma_\theta=0.08$, $\sigma_y=0.15$, and initial distribution
\[
  \theta_0\sim N(0,0.2^2).
\]
Although the data-generating trajectory is deterministic, the comparison is conditional on this filtering model: the Kalman filter gives the exact filtering distribution under the assumed random-walk model, while the particle filter approximates the same distribution.

Let
\[
  \theta_t\mid y_{1:t}\sim N(m_t,s_t^2)
\]
denote the Kalman filtering distribution. The exact posterior probability of being in the acceptable region is then
\[
  p_{A,t}
  =
  P(\theta_t\in A\mid y_{1:t})
  =
  \Phi\left(\frac{0.5-m_t}{s_t}\right)
  -
  \Phi\left(\frac{-0.5-m_t}{s_t}\right).
\]
The particle filter was run with $P=500$, $2000$, and $5000$ particles, using systematic resampling when the effective sample size fell below $0.5P$. Results are based on 200 Monte Carlo replications.

\begin{table}[tbp]
\centering
\scriptsize
\setlength{\tabcolsep}{4pt}
\begin{tabular*}{\hsize}{@{\extracolsep{\fill}}rrrrrrrr@{}}
\hline
$P$ & RMSE mean & RMSE $p_A$ & MAE $p_A$ & 95\% abs. err. $p_A$ & Max abs. err. $p_A$ & Mean ESS & Resamp. \\
\hline
500  & 0.0056 & 0.0102 & 0.0035 & 0.0221 & 0.2178 & 303.2  & 0.311 \\
2000 & 0.0028 & 0.0049 & 0.0017 & 0.0109 & 0.1213 & 1211.2 & 0.314 \\
5000 & 0.0018 & 0.0032 & 0.0011 & 0.0070 & 0.0687 & 3027.4 & 0.312 \\
\hline
\end{tabular*}
\caption{Particle-filter accuracy in the Gaussian tracking example. Errors are computed relative to the exact Kalman filtering distribution over 200 simulated sequences of length 200. Here $p_A=P(\theta_t\in A\mid y_{1:t})$, ESS denotes the effective sample size, and ``Resamp.'' is the proportion of time points at which resampling was performed.}
\label{tab:gaussian-tracking}
\end{table}

The particle approximation improves monotonically with the number of particles. With $P=5000$, the RMSE of the posterior acceptable-region probability is $0.0032$, and 95\% of absolute errors in this probability are below $0.0070$. Thus, in a setting where the exact answer is known, the particle filter accurately recovers both the posterior mean and the posterior probability of acceptable operation.

Using $\delta=0.5$ as a diagnostic threshold, none of the particle-filter runs missed the upward crossing of the acceptable boundary or the subsequent recovery. The mean signal delay after $t=78$ was approximately 1.45 observations for $P=5000$, and the mean recovery delay after $t=158$ was approximately 1.56 observations. These diagnostic delays are secondary; the main purpose of this example is to validate the particle approximation to $p_{A,t}$.


\subsection{Tracking a drifting Binomial proportion}
\label{subsec:binomial-tracking}

We next consider a nonlinear and non-Gaussian tracking problem. The latent parameter is a defect probability $\theta_t$, and the process is regarded as acceptable while
\[
  \theta_t \leq 0.02.
\]
At each time point a batch of $N=500$ items is inspected and
\[
  Y_t\mid \theta_t\sim \operatorname{Binomial}(N,\theta_t).
\]

The true latent probability follows a gradual degradation and recovery path:
\[
\theta_t =
\begin{cases}
0.01, & 1 \leq t \leq 50,\\
0.01 + 0.03(t-50)/60, & 51 \leq t \leq 110,\\
0.04, & 111 \leq t \leq 150,\\
0.04 - 0.03(t-150)/50, & 151 \leq t \leq 200.
\end{cases}
\]
With the convention that the process is out of control when $\theta_t>0.02$, the first unacceptable time point is $t=71$. The process returns to the acceptable region at $t=184$.

The filtering model tracks the logit-transformed state
\[
  z_t=\log\{\theta_t/(1-\theta_t)\}.
\]
A random-walk evolution is used:
\[
  z_t\mid z_{t-1}\sim N(z_{t-1},\sigma_z^2),
  \qquad \sigma_z=0.08,
\]
with observation model
\[
  Y_t\mid z_t\sim \operatorname{Binomial}
  \{N,\operatorname{logit}^{-1}(z_t)\}.
\]
The acceptable region on the logit scale is
\[
  z_t\leq \operatorname{logit}(0.02).
\]

For each replicate, a Phase I sample of $m_I=50$ in-control batches is generated from $\operatorname{Binomial}(500,0.01)$. Starting from the prior $\theta_0\sim\operatorname{Beta}(1,99)$, the Phase I posterior is
\[
  \theta_0\mid X_{1:m_I}
  \sim
  \operatorname{Beta}
  \left(
  1+\sum_{i=1}^{m_I}X_i,\,
  99+m_I N-\sum_{i=1}^{m_I}X_i
  \right).
\]
Initial particles are drawn from this posterior and transformed to the logit scale.

The threshold $\delta$ is chosen by posterior-predictive calibration, using the same finite-horizon false-signal criterion as in Section~\ref{subsec:threshold-calibration}. For each calibration replicate, Phase I data are generated, initial particles are drawn from the corresponding Phase I posterior, and an all-in-control Phase II sequence is generated from a rate drawn from that posterior. Candidate thresholds are evaluated from the stored posterior paths. Calibration and evaluation were performed separately for $P=1000$ and $P=5000$ particles.

\begin{table}[tbp]
\centering
\scriptsize
\setlength{\tabcolsep}{4pt}
\begin{tabular*}{\hsize}{@{\extracolsep{\fill}}rrrrrrrrr@{}}
\hline
$P$ & $\delta$ & Cal. $F$ & $d_1$ & $d_2$ & Miss & $F_{\rm pre}$ & Mean ESS & Resamp. \\
\hline
1000 & 0.990 & 1.24 (0.12) & 0.01 (0.00) & 12.65 (0.11) & 0.000/0.082 & 1.66 (0.04) & 669.0 & 0.186 \\
5000 & 0.985 & 0.79 (0.08) & 0.03 (0.01) & 12.09 (0.11) & 0.000/0.048 & 1.59 (0.04) & 3345.3 & 0.185 \\
\hline
\end{tabular*}
\caption{Sequential tracking of a drifting Binomial defect probability. The process is unacceptable when $\theta_t>0.02$. The threshold $\delta$ is calibrated separately for each particle count. Here $d_1$ is the delay after the first unacceptable time point $t=71$, $d_2$ is the delay after return to the acceptable region at $t=184$, and $F_{\rm pre}$ is the mean number of signal episodes beginning before $t=71$. Monte Carlo standard errors are shown in parentheses for calibrated false-signal episodes, delays, and $F_{\rm pre}$.}
\label{tab:binomial-tracking}
\end{table}

The calibrated thresholds are close to one because, under in-control operation, the posterior probability of the acceptable event $\theta_t\leq 0.02$ is usually very high. Thus a high posterior threshold is needed to produce approximately one false-signal episode over 200 in-control observations. This does not mean that the posterior probability of being unacceptable is near one at signalling;
rather, it reflects the calibration target for a process whose in-control posterior acceptability
probability is usually extremely close to one. With these thresholds, the first unacceptable crossing is detected essentially immediately: the mean delay is $0.006$ for $P=1000$ and $0.032$ for $P=5000$, with no missed detections. Recovery is slower, with mean delay about 12 observations and miss rates of 0.082 and 0.048 respectively. The pre-crossing signal rate is higher than the all-in-control calibration target because the latent process is already drifting toward the unacceptable boundary before $t=71$.

The effective sample size diagnostics are stable. Mean ESS is approximately $0.67P$ for both particle counts, and resampling occurs at about 18.5\% of time points. No particle-filter failures or numerical probability corrections occurred in these runs.

As a threshold sensitivity check, fixed thresholds $\delta=0.5$ and $\delta=0.1$ were also applied to the same posterior paths. For $P=5000$, using $\delta=0.5$ increased the first detection delay to 2.93 observations but reduced pre-crossing signal episodes to 0.258, while using $\delta=0.1$ increased the first detection delay to 8.97 observations and reduced pre-crossing signal episodes to 0.006. This illustrates the same trade-off seen in Section~\ref{sec:simulation}: posterior probabilities provide the monitoring statistic, while the threshold controls the operating behaviour.


\section{Multivariate data illustration: white wine quality}
\label{sec:wine-illustration}

We finally give a held-out multivariate illustration of the sequential-tracking method using the white wine quality data considered by \cite{Zou2011} and \cite{Tan2012}. The data contain 4898 observations, each consisting of 11 physicochemical measurements and a sensory quality score. The observations are not naturally time ordered, so the analysis below should not be interpreted as a case study of a naturally observed production stream. Instead, we use the data to construct pseudo-monitoring sequences from disjoint held-out pools of acceptable and degraded observations. The purpose is deliberately narrower than a full industrial case study: it illustrates the multivariate acceptable-region construction, posterior threshold calibration, and sequential propagation of uncertainty in a setting with real physicochemical measurements.

As in the earlier studies using these data, wines with quality score 7 are treated as acceptable production and wines with quality score 6 are treated as degraded production. Other quality levels are not used in this illustration. The quality-7 observations are split into three disjoint groups: 440 reference observations, 220 calibration observations, and 220 test observations. The 2198 quality-6 observations are used only as degraded test observations. All 11 physicochemical variables are standardised using the means and standard deviations of the quality-7 reference set. The quality-6 data are not used in standardisation, covariance estimation, definition of the acceptable region, or threshold calibration.

Let \(Y_t\in\mathbb{R}^{11}\) denote the standardised vector of wine measurements at time \(t\), and let \(\theta_t\) denote the latent current mean vector. We use the Gaussian tracking model
\[
  Y_t\mid \theta_t \sim N_{11}(\theta_t,\widehat\Sigma_y),
  \qquad
  \theta_t\mid\theta_{t-1}\sim
  N_{11}\left(\theta_{t-1},c_\theta \widehat\Sigma_y/b\right).
\]
Here \(\widehat\Sigma_y\) is a regularised covariance matrix estimated from the quality-7 reference set. Specifically, if \(\widehat\Sigma_{\rm raw}\) denotes the sample covariance matrix after standardisation, then
\[
  \widehat\Sigma_y
  =
  (1-\rho)\widehat\Sigma_{\rm raw}
  +
  \rho\,\operatorname{diag}\{\operatorname{diag}(\widehat\Sigma_{\rm raw})\},
  \qquad
  \rho=0.05.
\]
This reduced the condition number from 359.7 for the raw covariance matrix to 60.0. The block size \(b=10\) is used because the acceptable region is intended to describe a latent short-run mean, rather than an individual noisy wine observation. Initial particles are drawn from
\[
  \theta_0\sim N_{11}(\widehat\mu_{\rm ref},\widehat\Sigma_y/b),
\]
where \(\widehat\mu_{\rm ref}\) is the reference mean vector, which is approximately zero after standardisation. In the results below we use \(c_\theta=0.20\) and \(P=5000\) particles.

The acceptable region is an ellipsoid for the latent mean vector:
\[
  A =
  \left\{
  \theta:
  (\theta-\widehat\mu_{\rm ref})^\top
  \widehat\Sigma_y^{-1}
  (\theta-\widehat\mu_{\rm ref})
  \leq c_A
  \right\}.
\]
The threshold \(c_A\) is chosen from the quality-7 calibration set. We generate 5000 bootstrap means of blocks of \(b=10\) calibration observations and compute their squared Mahalanobis distances from \(\widehat\mu_{\rm ref}\). The value \(c_A\) is then set to the empirical 95th percentile of these distances, giving
\[
  c_A = 2.623.
\]
As a static check, the same block-mean Mahalanobis distance was computed for held-out quality-7 test observations and for quality-6 observations. Only 1.2\% of quality-7 test block means fell outside \(A\), compared with 37.0\% of quality-6 block means. Thus the full 11-dimensional measurements contain a moderate but visible shift between quality-7 and quality-6 wines under this latent-mean region.

\begin{figure}[tbp]
\centering
\includegraphics[width=0.75\textwidth]{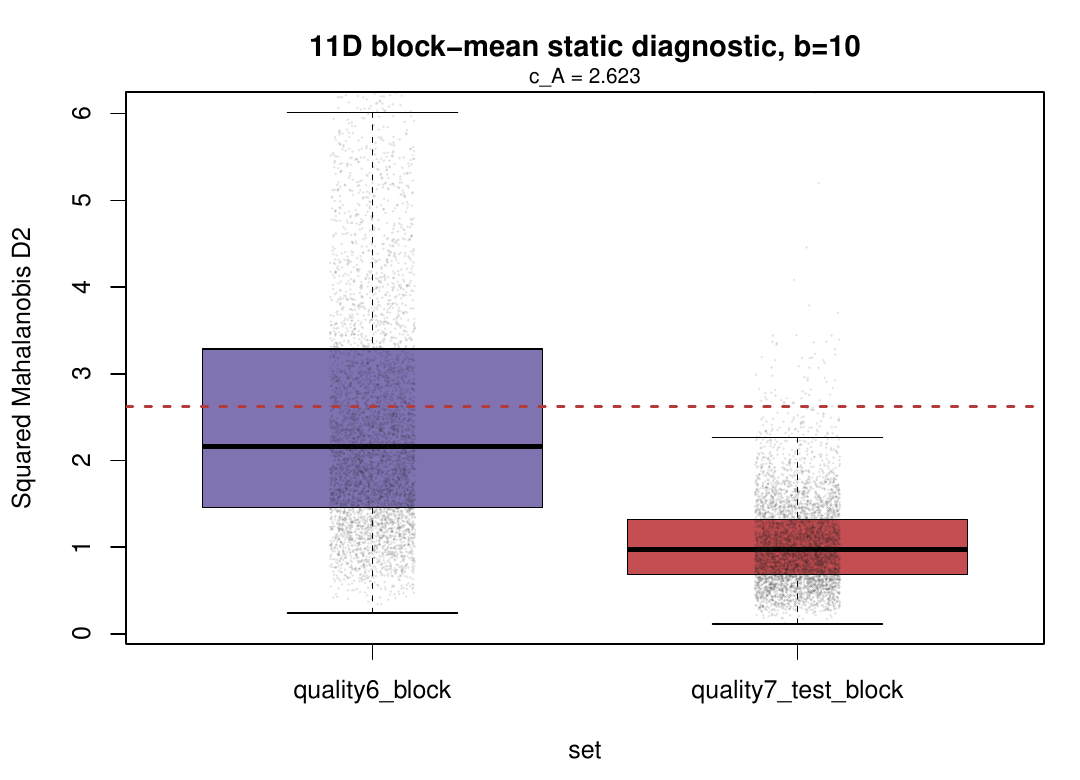}
\caption{Static diagnostic for the wine quality illustration. The plot shows squared Mahalanobis distances for bootstrap block means of held-out quality-7 observations and quality-6 observations, using the 11-dimensional latent-mean acceptable region. The dashed line is the acceptable-region threshold \(c_A=2.623\), determined from quality-7 calibration block means.}
\label{fig:wine-mahalanobis}
\end{figure}

The posterior signalling threshold is calibrated using all-quality-7 pseudo-sequences. Each calibration sequence consists of 150 observations sampled with replacement from the quality-7 calibration and test pools. For each sequence, the particle filter is run once and the posterior path
\[
  p_{A,t}=P(\theta_t\in A\mid y_{1:t})
\]
is stored. Candidate values of \(\delta\) are then evaluated from these stored paths. To avoid calibrations that achieve a target number of signal episodes only by remaining in signal for long periods, we restrict attention to thresholds producing between 0.75 and 1.25 false-signal episodes per 150 all-quality-7 observations and then choose the threshold with the smallest mean number of false-signal timepoints. With 500 calibration sequences, this gives
\[
  \delta=0.205.
\]
The achieved mean number of false-signal episodes is 0.758, with Monte Carlo standard error 0.046, and the mean number of false-signal timepoints is 1.550, with Monte Carlo standard error 0.115.

For evaluation, we construct pseudo-monitoring sequences of length 150:
\[
  50\ \text{quality-7 observations}
  \;\rightarrow\;
  50\ \text{quality-6 observations}
  \;\rightarrow\;
  50\ \text{quality-7 observations}.
\]
Observations are sampled with replacement from the corresponding held-out pools. A detection is defined as a new signal episode beginning during the quality-6 segment. If the chart is already signalling before the quality-6 segment begins, this is counted as false signalling in the initial quality-7 segment rather than as a successful detection. Recovery is defined as the first return to non-signal during the final quality-7 segment, conditional on a detection having occurred.

\begin{table}[tbp]
\centering
\scriptsize
\setlength{\tabcolsep}{4pt}
\begin{tabular*}{\hsize}{@{\extracolsep{\fill}}lcc@{}}
\hline
Quantity & Mean / rate & MCSE \\
\hline
Detection delay \(d_1\) & 16.60 & 0.36 \\
Detection miss rate & 0.031 & -- \\
Recovery delay \(d_2\) & 1.62 & 0.05 \\
Recovery miss rate & 0.000 & -- \\
False-signal episodes in initial quality-7 segment & 0.133 & 0.014 \\
False-signal timepoints in initial quality-7 segment & 0.212 & 0.030 \\
Already signalling at \(t=50\) & 0.003 & -- \\
Mean ESS & 1597.6 & -- \\
Resampling proportion & 0.855 & -- \\
\hline
\end{tabular*}
\caption{Performance of the 11-dimensional wine pseudo-monitoring illustration over 1000 constructed sequences. The sequence structure is 50 quality-7 observations, followed by 50 quality-6 observations, followed by 50 quality-7 observations. Detection delay \(d_1\) is measured from the start of the quality-6 segment and is based on a new signal episode beginning in that segment. Recovery delay \(d_2\) is measured from the start of the final quality-7 segment.}
\label{tab:wine-results}
\end{table}

The method detects the degraded quality-6 segment in most sequences. The mean detection delay is 16.60 observations, with a miss rate of 0.031 over the 50-observation degraded segment. Recovery is rapid once the sequence returns to quality-7 observations: the mean recovery delay is 1.62 observations and no recovery misses occur among detected sequences. False signalling in the initial quality-7 segment is low, with 0.133 false-signal episodes and 0.212 false-signal timepoints on average. The rate of already being in signal at the transition to quality 6 is only 0.003. Effective sample size diagnostics are stable, although resampling is frequent in this 11-dimensional example.

\begin{figure}[tbp]
\centering
\includegraphics[width=0.75\textwidth]{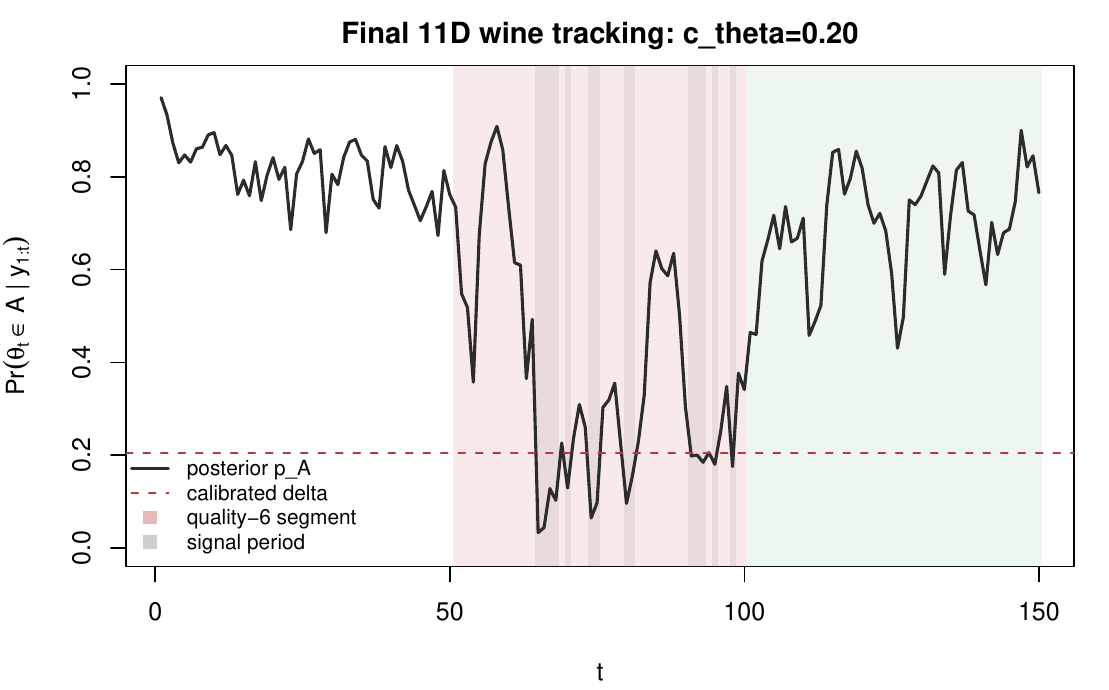}
\caption{Representative posterior acceptability path for the wine pseudo-monitoring illustration. The shaded middle region corresponds to the quality-6 segment. The solid curve shows \(P(\theta_t\in A\mid y_{1:t})\), the horizontal line is the calibrated threshold \(\delta=0.205\), and the signal period indicates times at which the posterior acceptability probability is below the threshold.}
\label{fig:wine-tracking}
\end{figure}

This example should be interpreted as a held-out multivariate illustration of the monitoring statistic rather than as a benchmark study of wine-quality classification or a naturally observed production stream. The quality-6 and quality-7 classes overlap, and the pseudo-sequences are constructed by sampling observations from quality classes rather than by observing a process through time. Nevertheless, the example demonstrates how the proposed sequential-tracking framework can be applied in a multivariate setting with a data-derived acceptable region, a calibrated posterior threshold, and full propagation of uncertainty in the latent current mean.

\section{Conclusion}

This paper has treated Phase II process monitoring as a problem of sequential
inference on current process acceptability. In many applications the practical
question is not only whether a process has changed at some point in the past,
but whether it is presently operating acceptably. This perspective is especially
useful when the in-control distribution must be learned from limited Phase I
data, when monitoring continues after a signal, when corrective action may lead
to recovery, or when a time-varying parameter is allowed to vary within a
prespecified acceptable region.

We developed sequential Bayesian methods for two instances of this
current-status problem. First, we considered recoverable processes which may
move from an in-control state to an out-of-control state, subsequently return to
in-control operation, and later degrade again. For this setting we formulated an
explicit recurrent in-control/out-of-control state model and derived recursive
updates for the posterior probability that the process is currently in control.
This posterior probability provides a natural monitoring statistic for applications
where the process cannot simply be stopped after the first signal.

Second, we considered sequential tracking problems in which the process
parameter is allowed to evolve over time and the aim is to monitor whether it
remains inside a prespecified acceptable region. This formulation separates the
region of practically acceptable parameter values from the posterior signalling
threshold. In linear Gaussian cases the relevant posterior probabilities can be
computed analytically, while in nonlinear, non-Gaussian, or multivariate settings
they can be approximated using Sequential Monte Carlo methods.

A recurring theme throughout the paper is that posterior probabilities are not
by themselves operating characteristics. A posterior threshold such as
$\delta=0.5$ or $\delta=0.9$ is interpretable as a decision rule only after the
losses or desired false-signal behaviour have been specified. We therefore used
posterior-predictive calibration to choose signalling thresholds with specified
finite-horizon in-control behaviour. This calibration propagates uncertainty in
the in-control parameter rather than treating a Phase I estimate as if it were the
true parameter value.

The operating-characteristic study illustrates both the strengths and the limitations of the
approach. In the Exponential--Gamma time-between-failure setting, the
recoverable-regime filter gave competitive false-signal behaviour and shorter
mean detection and recovery delays than a classical exponential CPM comparator.
The sensitivity analyses also show that the method is not prior-free. In
particular, the prior for out-of-control segment parameters affects which
departures are detected efficiently, and very limited Phase I information can
lead to unstable behaviour unless the in-control prior is informative. This sensitivity reflects the modelling problem rather than the numerical computation:
an informed Bayesian chart is most useful when the prior represents the departures that
the practitioner wants to detect.

The sequential-tracking examples show how the same posterior acceptability
principle can be used outside the exact conjugate setting. In a Gaussian tracking
problem, the particle filter closely approximated the exact Kalman filtering
probabilities. In a Binomial-logit tracking problem, it provided a direct way to
monitor the posterior probability that a drifting defect probability remained
below an acceptable threshold. Finally, the white wine illustration showed how
the method can be applied to multivariate process measurements using a
data-derived acceptable region and a calibrated posterior threshold, while also
making clear the narrower role of a pseudo-monitoring example constructed
from data that are not naturally time ordered.

Overall, the results suggest that Bayesian filtering provides a useful framework
for Phase II monitoring when current process status, parameter uncertainty, and
practical acceptability are central concerns. The approach is not a replacement
for all existing SPC methodology, but it gives a coherent way to combine Phase I
uncertainty, sequential updating, recoverable process behaviour, and
application-specific acceptable regions within a single posterior monitoring
framework.

\section*{Data Availability Statement}

The white wine quality data used in Section 8 are publicly available from the
UCI Machine Learning Repository as the Wine Quality data set. The data set was
donated by Cortez, Cerdeira, Almeida, Matos and Reis (2009) and is available at
\url{https://archive.ics.uci.edu/dataset/186/wine+quality}. The analysis in this
paper uses the file \texttt{winequality-white.csv}. Simulation data are generated
from the models described in the paper.

\bibliographystyle{apa}
\bibliography{centralbibliography,zotero,MyLibrary}

@Article{Apley2012,
  Title                    = {Posterior Distribution Charts: A {B}ayesian Approach for Graphically Exploring a Process Mean},
  Author                   = {Daniel W. Apley},
  Journal                  = {Technometrics},
  Year                     = {2012},
  Number                   = {3},
  Pages                    = {279-293},
  Volume                   = {54}
}

@Article{Bayarri2005,
  Title                    = {A {B}ayesian Sequential Look at u-Control Charts},
  Author                   = {M. J. Bayarri and G. Garcia-Donato},
  Journal                  = {Technometrics},
  Year                     = {2005},
  Number                   = {2},
  Pages                    = {142-151},
  Volume                   = {47}
}

@InProceedings{Bodenham2013,
  Title                    = {Continuous monitoring of a computer network using multivariate adaptive estimation},
  Author                   = {Dean Bodenham and Niall Adams},
  Booktitle                = {{IEEE} Data Mining Workshops ({ICDMW})},
  Year                     = {2013},
  Pages                    = {311-318}
}

@Article{Chopin2007,
  Title                    = {Dynamic detection of change points in long time series},
  Author                   = {Chopin, Nicolas},
  Journal                  = {Annals of the Institute of Statistical Mathematics},
  Year                     = {2007},

  Month                    = {JUN},
  Number                   = {2},
  Pages                    = {349-366},
  Volume                   = {59},

  Doi                      = {10.1007/s10463-006-0053-9},
  ISSN                     = {0020-3157},
  Unique-id                = {ISI:000247152200008}
}

@Article{Crisan2002,
  Title                    = {A Survey of Convergence Results on Particle Filtering Methods for Practitioners},
  Author                   = {Dan Crisan and Arnaud Doucet},
  Journal                  = {IEEE Transactions on Signal Processing},
  Year                     = {2002},
  Number                   = {3},
  Pages                    = {736-746},
  Volume                   = {50}
}

@Book{Doucet2001,
  Title                    = {Sequential Monte Carlo Methods in Practice},
  Author                   = {Arnaud Doucet and Nando de Freitas and Neil Gordon},
  Publisher                = {Springer},
  Year                     = {2001}
}

@Book{Durbin2001,
  Title                    = {Time Series Analysis by State Space Methods},
  Author                   = {James Durbin and Siem Jan Koopman},
  Publisher                = {Oxford Statistical Science Series},
  Year                     = {2001}
}

@Article{Fearnhead2006,
  Title                    = {Exact and efficient {B}ayesian inference for multiple changepoint problems},
  Author                   = {Fearnhead, P},
  Journal                  = {Statistics and Computing},
  Year                     = {2006},
  Pages                    = {203-213},
  Volume                   = {16}
}

@Article{Fearnhead2007,
  Title                    = {On-line inference for multiple changepoint problems},
  Author                   = {Fearnhead, Paul and Liu, Zhen},
  Journal                  = {Journal of the Royal Statistical Society Series B},
  Year                     = {2007},
  Number                   = {4},
  Pages                    = {589-605},
  Volume                   = {69},
  ISSN                     = {1369-7412},
  Unique-id                = {ISI:000249250000004}
}

@Article{Gandy2013,
  Title                    = {Non-restarting {CUSUM} charts and control of the false discovery rate.},
  Author                   = {Axel Gandy and F. Din-Houn Lau},
  Journal                  = {Biometrika},
  Year                     = {2013},
  Volume                   = {100}
}

@Article{Jensen2006,
  Title                    = {Effects of Parameter Estimation on Control Chart Properties: A Literature Review},
  Author                   = {Jensen, Willis A. and Jones-Farmer, L. Allison and Champ, Charles W. and Woodall, William H.},
  Journal                  = {Journal of Quality Technology},
  Year                     = {2006},
  Number                   = {4},
  Pages                    = {349-364},
  Volume                   = {38},
  Unique-id                = {ISI:000241136100006}
}

@Article{Johansen2008,
  Title                    = {A note on auxiliary particle filters.},
  Author                   = {Johansen, A.M. and Doucet, A},
  Journal                  = {Statistics and Probability Letters},
  Year                     = {2008},
  Number                   = {12},
  Pages                    = {1498-1504},
  Volume                   = {78}
}

@Article{Jones2001,
  Title                    = {The Performance of Exponentially Weighted Moving Average Charts with Estimated Parameters},
  Author                   = {Jones, LA and Champ, CW and Rigdon, SE},
  Journal                  = {Technometrics},
  Year                     = {2001},
  Number                   = {2},
  Pages                    = {156-167},
  Volume                   = {43},
  ISSN                     = {0040-1706},
  Unique-id                = {ISI:000168205900005}
}

@Book{Montgomery2005,
  Title                    = {Introduction to Statistical Quality Control},
  Author                   = {D. C. Montgomery},
  Publisher                = {Wiley},
  Year                     = {2005}
}

@Book{Oakland2007,
  Title                    = {Statistical Process Control},
  Author                   = {John Oakland},
  Publisher                = {Routledge},
  Year                     = {2007}
}

@Article{Pitt1999,
  Title                    = {Filtering via simulation: auxiliary particle filter},
  Author                   = {Michael Pitt and Neil Shephard},
  Journal                  = {Journal of the American Statistical Association},
  Year                     = {1999},
  Pages                    = {590-599},
  Volume                   = {94}
}

@Article{RossSC,
  Title                    = {Sequential Change Detection in the Presence of Unknown Parameters},
  Author                   = {Gordon J. Ross},
  Journal                  = {Statistics and Computing},
  Year                     = {2014},
  Pages                    = {1017-1030},
  Volume                   = {24}
}

@Article{Rosscompstat,
  Title                    = {A Change Point Model for Monitoring a Proportion},
  Author                   = {Gordon J. Ross and Dimitris K. Tasoulis and Niall M. Adams},
  Journal                  = {Computational Statistics},
  Year                     = {2013},
  Number                   = {2},
  Pages                    = {463-479},
  Volume                   = {28}
}

@Article{Rosstechnometrics,
  Title                    = {Nonparametric Monitoring of Data Streams for Changes in Location and Scale},
  Author                   = {Gordon J. Ross and Dimitris K. Tasoulis and Niall M. Adams},
  Journal                  = {Technometrics},
  Year                     = {2011},
  Number                   = {4},
  Pages                    = {379-389},
  Volume                   = {53}
}

@Article{Spiegelhalter2012,
  Title                    = {Statistical methods for healthcare regulation: rating, screening and surveillance},
  Author                   = {David Spiegelhalter and Christopher Sherlaw-Johnson and Martin Bardsley and Ian Blunt and Christopher Wood Olivia Grigg},
  Journal                  = {Journal of the Royal Statistical Society: Series A},
  Year                     = {2012},
  Number                   = {1},
  Pages                    = {1-47},
  Volume                   = {175}
}

@Article{Tan2012,
  Title                    = {A {B}ayesian Approach for Interpreting Mean Shifts in Multivariate Quality Control},
  Author                   = {Matthias H. Y. Tan and Jianjun Shi},
  Journal                  = {Technometrics},
  Year                     = {2012},
  Number                   = {3},
  Pages                    = {294-307},
  Volume                   = {54}
}

@Article{Tsiamyrtzis2010,
  Title                    = {{B}ayesian Startup Phase Mean Monitoring of an Autocorrelated Process That Is Subject to Random Sized Jumps},
  Author                   = {Panagiotis Tsiamyrtzis and Douglas M. Hawkins},
  Journal                  = {Technometrics},
  Year                     = {2010},
  Number                   = {4},
  Pages                    = {438-452},
  Volume                   = {52}
}

@Article{Tsiamyrtzis2008,
  Title                    = {A {B}ayesian {EWMA} method to detect jumps at the start-up phase of a process},
  Author                   = {Panagiotis Tsiamyrtzis and Douglas M. Hawkins},
  Journal                  = {Quality and Reliability Engineering International},
  Year                     = {2008},
  Number                   = {6},
  Pages                    = {721-735},
  Volume                   = {24}
}

@Article{Tsiamyrtzis2005,
  Title                    = {A {B}ayesian Scheme to Detect Changes in the Mean of a Short-Run Process},
  Author                   = {Panagiotis Tsiamyrtzis and Douglas M. Hawkins},
  Journal                  = {Technometrics},
  Year                     = {2005},
  Number                   = {4},
  Pages                    = {446-457},
  Volume                   = {47}
}

@Article{Woodall1997,
  Title                    = {Control charts based on attribute data: Bibliography and review},
  Author                   = {Woodall, WH},
  Journal                  = {Journal of Quality Technology},
  Year                     = {1997},
  Number                   = {2},
  Pages                    = {172-183},
  Volume                   = {29}
}

@Article{Yeh2004,
  Title                    = {A likelihood-ratio-based {EWMA} control chart for monitoring variability of multivariate normal processes},
  Author                   = {Yeh, AB and Huwang, LC and Wu, YF},
  Journal                  = {IIE Transactions},
  Year                     = {2004},
  Number                   = {9},
  Pages                    = {865-879},
  Volume                   = {36},

  Doi                      = {10.1080/07408170490473042},
  ISSN                     = {0740-817X},
  Unique-id                = {ISI:000223349100007}
}

@Article{Zou2011,
  Title                    = {A {LASSO}-Based Diagnostic Framework for Multivariate Statistical Process Control,},
  Author                   = {Zou, C and Jiang, W and Tsung, F},
  Journal                  = {Technometrics},
  Year                     = {2011},
  Pages                    = {297-309},
  Volume                   = {53}
}

@article{Kumar2017,
author = {Nirpeksh Kumar and Subha Chakraborti},
title = {Bayesian Monitoring of Times Between Events: The Shewhart tr-Chart},
journal = {Journal of Quality Technology},
volume = {49},
number = {2},
pages = {136-154},
year  = {2017},
publisher = {Taylor & Francis},
doi = {10.1080/00224065.2017.11917985},

URL = { 
        https://doi.org/10.1080/00224065.2017.11917985
    
},
eprint = { 
        https://doi.org/10.1080/00224065.2017.11917985
    
}
}

@article{Pan2017,
author = {Rong Pan and Steven E. Rigdon},
title = {A Bayesian Approach to Change Point Estimation in Multivariate SPC},
journal = {Journal of Quality Technology},
volume = {44},
number = {3},
pages = {231-248},
year  = {2012},
publisher = {Taylor & Francis},
doi = {10.1080/00224065.2012.11917897},

URL = { 
        https://doi.org/10.1080/00224065.2012.11917897
    
},
eprint = { 
        https://doi.org/10.1080/00224065.2012.11917897
    
}

}

@article{Bourazas2022PCC,
  author  = {Bourazas, Konstantinos and Kiagias, Dimitrios and Tsiamyrtzis, Panagiotis},
  title   = {Predictive Control Charts ({PCC}): A Bayesian approach in online monitoring of short runs},
  journal = {Journal of Quality Technology},
  year    = {2022},
  volume  = {54},
  number  = {4},
  pages   = {367--391},
  doi     = {10.1080/00224065.2021.1916413}
}

@article{Bourazas2023PRC,
  author  = {Bourazas, Konstantinos and Sobas, Fr{\'e}d{\'e}ric and Tsiamyrtzis, Panagiotis},
  title   = {Predictive ratio {CUSUM} ({PRC}): A Bayesian approach in online change point detection of short runs},
  journal = {Journal of Quality Technology},
  year    = {2023},
  volume  = {55},
  number  = {4},
  pages   = {391--403},
  doi     = {10.1080/00224065.2022.2161434}
}

@article{Bourazas2023PRCDesign,
  author  = {Bourazas, Konstantinos and Sobas, Fr{\'e}d{\'e}ric and Tsiamyrtzis, Panagiotis},
  title   = {Design and properties of the predictive ratio {CUSUM} ({PRC}) control charts},
  journal = {Journal of Quality Technology},
  year    = {2023},
  volume  = {55},
  number  = {4},
  pages   = {404--421},
  doi     = {10.1080/00224065.2022.2161435}
}

@article{Ali2020TBE,
  author  = {Ali, Sajid},
  title   = {A predictive Bayesian approach to sequential time-between-events monitoring},
  journal = {Quality and Reliability Engineering International},
  year    = {2020},
  volume  = {36},
  number  = {1},
  pages   = {365--387},
  doi     = {10.1002/qre.2580}
}

@article{CapizziMasarotto2020,
  author  = {Capizzi, Giovanna and Masarotto, Guido},
  title   = {Guaranteed in-control control chart performance with cautious parameter learning},
  journal = {Journal of Quality Technology},
  year    = {2020},
  volume  = {52},
  number  = {4},
  pages   = {385--403},
  doi     = {10.1080/00224065.2019.1640096}
}

@article{WangCastagliolaGuo2026,
  author  = {Wang, Cheng and Castagliola, Philippe and Guo, Baocai},
  title   = {On Designing the Phase {II} Bayesian {CUSUM} Control Charts for Monitoring the Process Mean Under Guaranteed In-Control Performance},
  journal = {Quality and Reliability Engineering International},
  year    = {2026},
  doi     = {10.1002/qre.70222},
  note    = {Early View}
}

@article{AdamsMacKay2007,
  author  = {Adams, Ryan Prescott and MacKay, David J. C.},
  title   = {Bayesian Online Changepoint Detection},
  journal = {arXiv preprint arXiv:0710.3742},
  year    = {2007},
  doi     = {10.48550/arXiv.0710.3742}
}

@article{zhang_monitoring_2016,
	title = {Monitoring wafers’ geometric quality using an additive {Gaussian} process model},
	volume = {48},
	number = {1},
	journal = {IIE Transactions},
	author = {Zhang, Linmiao and Wang, Kaibo and Chen, Nan},
	year = {2016},
	pages = {1--15}
}

@article{chukhrova_improved_2019,
	title = {Improved control charts for fraction non-conforming based on hypergeometric distribution},
	volume = {128},
	urldate = {2021-06-21},
	journal = {Computers \& Industrial Engineering},
	author = {Chukhrova, Nataliya and Johannssen, Arne},
	year = {2019},
	pages = {795--806}
}

@article{bodenham_continuous_2017,
	title = {Continuous monitoring for changepoints in data streams using adaptive estimation},
	volume = {27},
	number = {5},
	urldate = {2021-06-21},
	journal = {Statistics and Computing},
	author = {Bodenham, Dean A. and Adams, Niall M.},
	year = {2017},
	pages = {1257--1270}
}

@article{wang_process_2018,
	title = {Process tracking and monitoring based on discrete jumping model},
	volume = {50},
	number = {1},
	journal = {Journal of Quality Technology},
	author = {Wang, Chao and Zhou, Shiyu},
	year = {2018},
	pages = {34--48}
}

@article{hou_new_2020,
	title = {A new {Bayesian} scheme for self-starting process mean monitoring},
	volume = {17},
	number = {6},
	journal = {Quality Technology \& Quantitative Management},
	author = {Hou, Yuxing and He, Baosheng and Zhang, Xudong and Chen, Yong and Yang, Qingyu},
	year = {2020},
	pages = {661--684}
}

@article{noor-ul-amin_adaptive_2021,
	title = {An adaptive {EWMA} control chart for monitoring the process mean in {Bayesian} theory under different loss functions},
	volume = {37},
	number = {2},
	journal = {Quality and Reliability Engineering International},
	author = {Noor-ul-Amin, Muhammad and Noor, Surria},
	year = {2021},
	pages = {804--819}
}

@article{yazdi_new_2019,
	title = {A new {Bayesian} multivariate exponentially weighted moving average control chart for phase {II} monitoring of multivariate multiple linear profiles},
	volume = {35},
	number = {7},
	journal = {Quality and Reliability Engineering International},
	author = {Yazdi, Ahmad Ahmadi and Hamadani, Ali Zeinal and Amiri, Amirhossein and Grzegorczyk, Marco},
	year = {2019},
	pages = {2152--2177}
}

@article{steward_bayesian_2016,
	title = {A {Bayesian} {Approach} to {Diagnostics} for {Multivariate} {Control} {Charts}},
	volume = {48},
	number = {4},
	journal = {Journal of Quality Technology},
	author = {Steward, Robert M. and Rigdon, Steven E. and Pan, Rong},
	year = {2016},
	pages = {303--325}
}

@article{zhao_alternating_2021,
	title = {Alternating {Pruned} {Dynamic} {Programming} for {Multiple} {Epidemic} {Change}-{Point} {Estimation}},
	volume = {30},
	number = {3},
	journal = {Journal of Computational and Graphical Statistics},
	author = {Zhao, Zifeng and Yau, Chun Yip},
	year = {2021},
	pages = {1--14}
}

@article{dette_detecting_2016,
	title = {Detecting relevant changes in time series models},
	volume = {78},
	number = {2},
	journal = {Journal of the Royal Statistical Society: Series B (Statistical Methodology)},
	author = {Dette, Holger and Wied, Dominik},
	year = {2016},
	pages = {371--394}
}

\end{document}